\documentclass[journal]{IEEEtran}

\usepackage{./bs_macros}
\usepackage{caption}
\usepackage[labelformat=simple]{subcaption}

\usepackage{float}
\usepackage{makecell}
\usepackage[noend]{algpseudocode}

\graphicspath{{./}}

\theoremstyle{definition}

\newtheorem{remark}{Remark}

\makeatletter
\patchcmd{\@maketitle}
{\addvspace{0.5\baselineskip}\egroup}
{\addvspace{-1\baselineskip}\egroup}
{}
{}
\makeatother

\begin{document}

\title{ \vspace{-0.5cm} \large  \textbf{Analysing Global Fixed Income Markets with Tensors} }

\author{
	
	\small Bruno Scalzo Dees
	
\thanks{B. Scalzo Dees is with the Department of Electrical Engineering, Imperial College London, London SW7 2AZ, U.K. (e-mail: bs1912@ic.ac.uk).}


}


\maketitle

\begin{abstract}
Global fixed income returns span across multiple maturities and economies, that is, they naturally reside on multi-dimensional data structures referred to as \textit{tensors}. In contrast to standard ``flat-view'' multivariate models that are agnostic to data structure and only describe linear pairwise relationships, we introduce a tensor-valued approach to model the global risks shared by multiple interest rate curves. In this way, the estimated risk factors can be analytically decomposed into \textit{maturity-domain} and \textit{country-domain} constituents, which allows the investor to devise rigorous and tractable global portfolio management and hedging strategies tailored to each risk domain. An empirical analysis confirms the existence of global risk factors shared by eight developed economies, and demonstrates their ability to compactly describe the global macroeconomic environment.
\end{abstract}


\IEEEpeerreviewmaketitle

\vspace{-0.5cm}

\section{Introduction}

Market participants have long recognized the importance of identifying the \textit{common factors} that affect the returns of securities within asset classes. In such a task it is critical to distinguish the \textit{common} risks that have general impact on the returns of most securities from the \textit{idiosyncratic} risks that influence securities individually.  For instance, following the seminal work in \cite{Litterman1991}, a significant portion of the fixed income literature has been devoted to the technique of principal component analysis (PCA) \cite{Jolliffe1986} to provide a parsimonious interpretation to the dynamics of the term structure of interest rates. Empirical results suggest that three latent factors, referred to as \textit{level}, \textit{slope} and \textit{curvature}, are required to almost fully reflect the behaviour of the entire term structure. Moreover, the principal components are frequently identified with economically meaningful events. As such, the degree of robustness in these findings has made PCA a fundamental building block for characterizing single-economy interest rate curves. Notable benefits of the principal components approach include: (i) its analyticity and mathematical tractability; (ii) its ability to parsimoniously describe economic factors; (iii) its applications for stress-testing and scenario analysis; (iv) its direct applicability for hedging portfolios. 

However, the growing interconnectedness of the international markets presents a major challenge for risk management of fixed-income securities, owing to the high correlation between interest rates across maturities and countries. Financial institutions routinely invest globally using strategies with limited avenues for diversification. This is largely due to the legacy analytics which employ ``flat-view'' multivariate methods, i.e. PCA, whereby trades are typically hedged by offsetting their domestic-curve principal components. This leaves such strategies unprotected to cross-country risk arising from global macroeconomic events. The most recent credit crisis, for instance, is exemplar of how macroeconomic shocks can be crucially transmitted across interest rate curves. For this reason, a parsimonious model to describe the co-variation of interest rates at the relevant maturities and in the relevant countries appears necessary for global fixed income investors to adequately identify and manage risk.

This has naturally motivated various developments of PCA for joint term structure analysis, however, a wide range of solutions have been proposed as a consequence of the ambiguity in the problem formulation. We find a substantial lack of agreement in the literature, not only regarding the data preparation and estimation procedure, but also regarding the number of latent factors that are required to explain the joint dynamics of multiple yield curves, and the nature of the global and domestic factors obtained. 

One approach is to apply PCA to vectorised data obtained from several term structures \cite{Rodrigues1997,Driessen2003,Novosyolov2008}, however, this method ignores the multi-curve structure and leads to factors which are difficult to interpret and can still reflect idiosyncratic and domestic behaviour. Another approach is \textit{common PCA} \cite{Flury1988}, which extracts the eigenvectors that span an identical space across all countries, however, this method simultaneously diagonalizes multiple covariance matrices, which is non-analytic for more than two matrices, and neglects the co-variation between assets across countries \cite{Juneja2012}. Alternatively, \textit{inter-battery factor analysis} \cite{Tucker1958} captures all common factors across domestic term structures \cite{Perignon2007}, however, the solution method is also computationally prohibitive and implicitly assumes that idiosyncratic co-variations can only occur domestically, which is a restrictive and unrealistic assumption. The major limitation shared by the existing techniques is that they are agnostic to the multi-curve data structure and resort to methods developed for multivariate analysis. Such a flattened view of the data, and the rigid assumptions inherent in multivariate analysis, are inadequate and ineffective.

We recognise that global fixed income returns, which span across multiple maturities and countries, reside on regular multi-dimensional data structures referred to as \textit{tensors}.  It is only through \textit{tensor analysis} that we have the opportunity to develop sophisticated models capturing the interactions between the entirety of interest rate curves. This motivates the development of \textit{multilinear} techniques, which have eventually found its place in many real-world applications where tensors naturally reside \cite{Kolda2009,Mandic2015_3,Mandic2017_2,Mandic2017_3,Sidiropoulos2017_1}.

Accordingly, we have developed a framework that employs the structure-aware \textit{multilinear algebra} to rigorously model the risk factors shared by an international universe of fixed income returns. In this way, the estimated risk factors can be analytically decomposed into two parallel domains of risk: (i) \textit{maturity-domain} factors which are shared by all countries; and (ii) \textit{country-domain} factors which are shared by all maturities. By operating within each domain in parallel, the investor can devise rigorous and tractable global portfolio management and hedging strategies, with fewer decision parameters, that are simultaneously tailored to each risk domain, as a consequence and natural extension of the proposed multilinear framework. An empirical analysis confirms the existence of common global risk factors shared by eight developed economies. The resulting maturity-domain and country-domain factors are shown to provide compact and physically meaningful insight into the global macroeconomic environment.

\newpage

\section{Prerequisites of Tensor Algebra}

\label{section:preliminaries}

Tensors are manipulated using the mathematical branch of \textit{multilinear algebra}, for which we provide a comprehensive introduction to the subject from an academically rigorous, yet practitioner-focused perspective. We refer the reader to \cite{Kolda2009,Mandic2015_3} for more details on the topic.

In this work, scalars are denoted by lightface font, e.g. $x$; vectors by lowercase boldface font, e.g. $\x$; matrices by uppercase boldface font, e.g. $\X$; and tensors by boldface calligraphic font, e.g. $\calX$.

\subsection{Nomenclature}

The \textit{order} of a tensor defines the number of dimensions, also referred to as \textit{modes}. For instance, the tensor  $\calX \in \domR^{I_{1}\times \cdots \times I_{N}}$ is of order $N$ and has $K = \prod_{n=1}^{N} I_{n}$ elements in total. 

Tensors can be reshaped into mathematically tractable vector and matrix representations, which we can manipulate using linear algebra. The \textit{vector representation} is denoted by 
\begin{align}
	\x = \vect{\calX} \quad \in \domR^{K}
\end{align}
and the \textit{mode-$n$ unfolding} is obtained by a reshaping the tensor into the matrix 
\begin{align}
	\X_{(n)}=\left[ \begin{array}{cccc}
		\f_{1}^{(n)} & \f_{2}^{(n)} & \cdots & \f_{\frac{K}{I_{n}}}^{(n)}
	\end{array} \right] \quad \in \domR^{I_{n} \times \frac{K}{I_{n}}}
\end{align}
which contains the set of column vectors, $ \f_{i}^{(n)} \in \domR^{I_{n}}$, known as \textit{mode-$n$ fibres}. Fibres are the multi-dimensional generalization of matrix rows and columns.

\begin{figure}[htbp]
	\centering
	\includegraphics[width=0.35\textwidth, trim={2cm 0 0 0},clip]{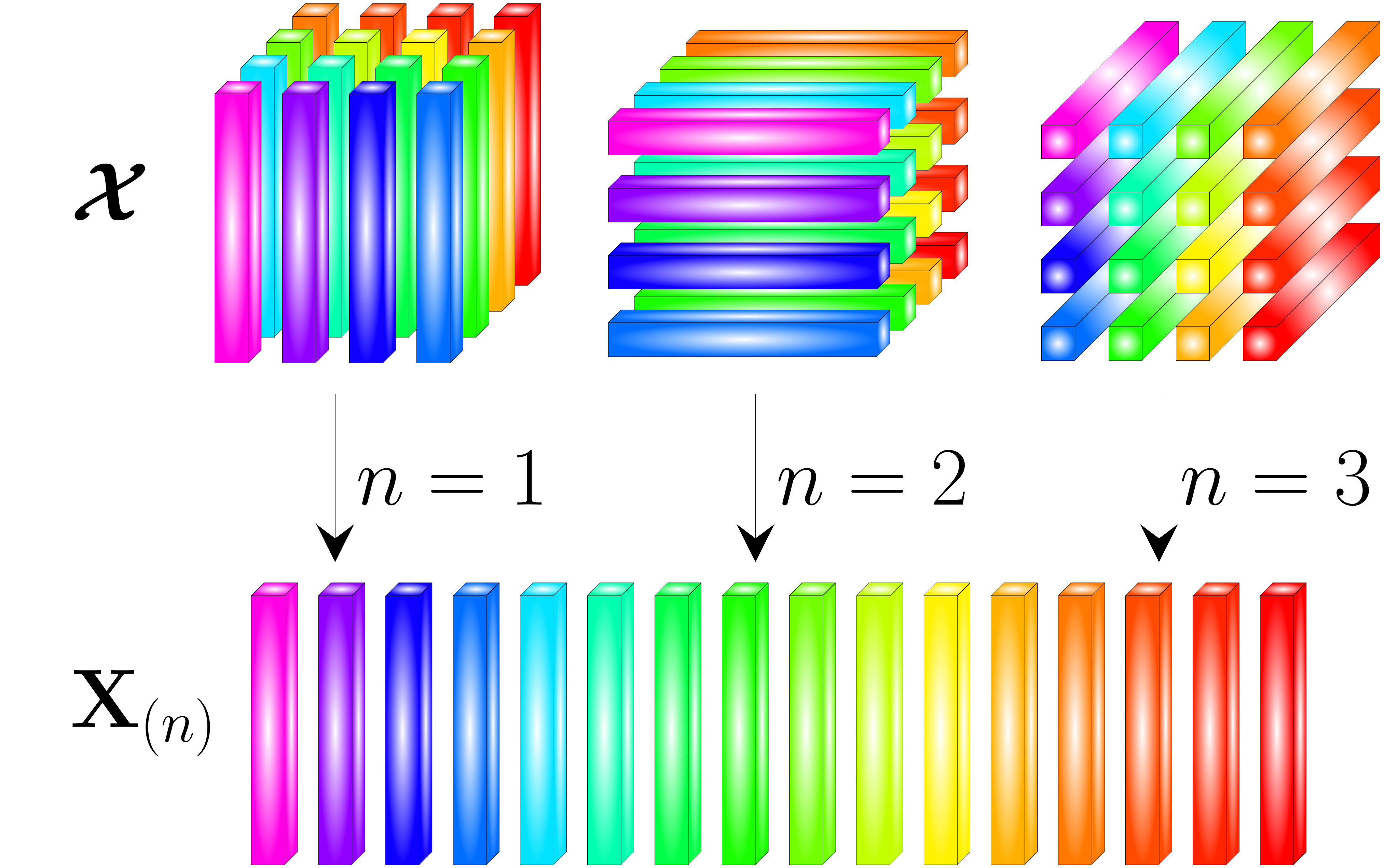}
	\caption{\label{fig:mode_n_unfold} Illustration of the mode-$n$ unfolding, $\X_{(n)}$, of the order-3 tensor, $\calX$, in view of the orientation of mode-$n$ fibres. }
\end{figure}

The operation of \textit{mode-$n$ unfolding} can be viewed as the reorientation of the mode-$n$ fibres as column vectors of $\X_{(n)}$, as illustrated in Figure \ref{fig:mode_n_unfold}. Notice that the order-$3$ tensor, $\calX$, has alternative, yet equivalent, representations in terms of mode-$1$ (left panel), mode-$2$ (middle panel) and mode-$3$ fibres (right panel), that is, columns, rows and tubes, respectively.

\subsection{Tensor products}

\begin{figure}[htbp]
	\centering
	\includegraphics[width=0.4\textwidth]{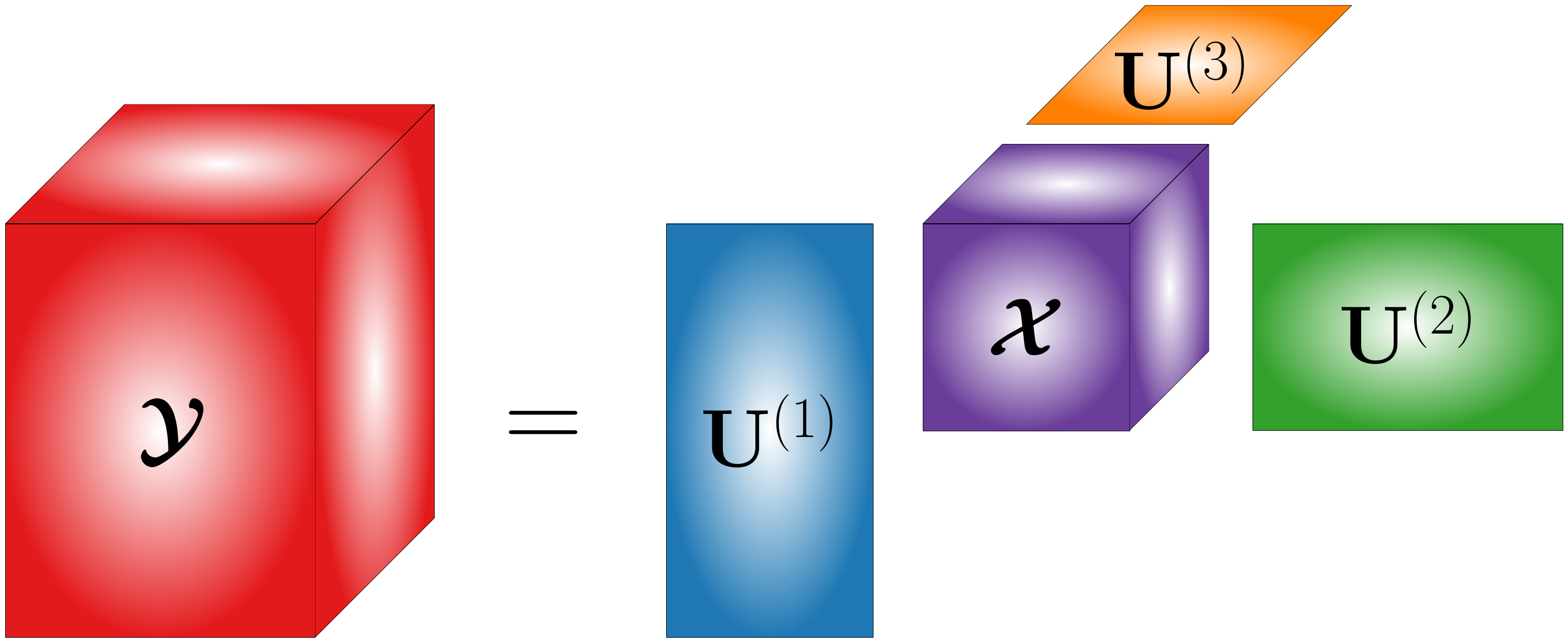}
	\caption{\label{fig:Tucker} Sequence of \textit{mode-$n$ products} for $n=1,2,3$.}
\end{figure}

Multilinear algebra is based on the class of operators known as \textit{tensor products}. The \textit{Kronecker product} between the matrices $\A \in \domR^{I\times I}$ and $\B \in \domR^{J \times J}$ yields the block matrix
\begin{align}
	\A \otimes \B = \left[\begin{array}{ccc}
		a_{11}\B & \cdots & a_{1I}\B \\
		\vdots & \ddots & \vdots \\
		a_{I1}\B & \cdots & a_{II}\B
	\end{array}\right] \quad \in \domR^{IJ \times IJ}
\end{align}
For convenience, we denote the sequence of Kronecker products of the matrices $\U^{(n)} \in \domR^{I_{n} \times I_{n}}$, for $n=1,...,N$, by
\begin{equation}
\left(\kronprod{n=1}{N}\U^{(n)}\right) = \U^{(1)}\otimes\cdots\otimes\U^{(N)} \quad \in \domR^{K \times K}
\end{equation}
The \textit{mode-$n$ product} of the tensor $\calX \in \domR^{I_{1}\times \cdots \times I_{N}}$ with the matrix $\U \in \domR^{J_{n} \times I_{n}}$ is denoted by 
\begin{align}
	\calY = \calX \times_{n} \U \quad \in \domR^{I_{1} \times \cdots \times I_{n-1} \times J_{n} \times I_{n+1} \times \cdots \times I_{N}}
\end{align}
and is equivalent to performing the following steps
\begin{spacing}{1.3}
	\begin{algorithmic}[1]
		\State $\X_{(n)} \leftarrow \calX $ \Comment{Mode-$n$ unfold}
		\State $\Y_{(n)} \leftarrow \U \X_{(n)}$ \Comment{Left matrix multiplication}
		\State $\calY \leftarrow \Y_{(n)}$ \Comment{Re-tensorize}
	\end{algorithmic}
\end{spacing}
\noindent For convenience, the sequence of \textit{mode-$n$ products} of $\calX$ and the matrices $\U^{(n)} \in \domR^{J_{n} \times I_{n}}$, for $n=1,...,N$, is written as
\begin{equation}
	\calY = \calX \modeprod{n=1}{N}\U^{(n)} = \calX \times_{1}\U^{(1)}\times_{2}\cdots\times_{N}\U^{(N)} \in \domR^{J_{1} \times \cdots \times J_{N}}
\end{equation}
and can be expressed in mathematically equivalent vector and matrix representations, that is
\begin{align}
	\y & = \left( \kronprod{\subalign{n&=N}}{1} \U^{(n)} \right)\x \\
	\Y_{(n)} & = \U^{(n)}\X_{(n)}\left( \kronprod{\subalign{i&=N\\i&\neq n}}{1} \U^{(i)\Trans} \right)
\end{align}
Figure \ref{fig:Tucker} illustrates the sequence of mode-$n$ products of an order-3 tensor with matrices $\U^{(n)}$, for $n=1,2,3$.

\subsection{Tensor-valued Gaussian random variables}

In standard multivariate data analysis, multiple measurements are collected at a given trial, experiment or time instant, to form the vector-valued sample, $\x \in \domR^{K}$. An assumption typically adopted in statistical modeling is that variables are described by the distribution $\x \sim \mathcal{N}\left( \m, \boldSigma \right)$, which implies that the covariance matrix, $\boldSigma \in \domR^{K \times K}$, is unstructured. However, if the variables have a natural tensor representation, then it is desirable, if not necessary, to assume that the covariance matrix, $\boldSigma$, exhibits a more structured form motivated by economic considerations. 

For instance, real-world order-$N$ tensor-valued signals encountered in finance include: 
\begin{enumerate}[label=\roman*)]
	\item interest rates over \textit{curve $\times$ maturity $\times$ country} ($N=3$);
	\item futures prices over \textit{asset $\times$ maturity} ($N=2$);
	\item options prices over \textit{asset $\times$ maturity $\times$ strike} ($N=3$).
\end{enumerate}

\noindent The statistical properties of tensor-valued random variables are intrinsically linked to that of Gaussian random fields, described as follows. Consider the zero-mean random variable on an $N$-dimensional coordinate system, denoted by $x : \domR^{N} \mapsto \domR$, and described by the coordinate-dependent distribution
\begin{align}
	x(\z) \sim \Normal{0,\sigma^{2}(\z)}
\end{align}
where $\z = \{ z^{(1)}, ..., z^{(N)} \} \in \domR^{N}$ is an $N$-dimensional coordinate vector, and $z^{(n)} \in \domR$ is the $n$-th axis coordinate. Furthermore, assume that the random variable is equipped with a covariance operator $\sigma : \domR^{N} \times \domR^{N} \mapsto \domR$
\begin{align}
	\sigma(\z_{1},\z_{2}) = \cov{x(\z_{1}),x(\z_{2})}
\end{align}
where $\sigma(\z,\z) \equiv \sigma^{2}(\z)$. The random variable is said to exhibit a \textit{separable} covariance structure if and only if the covariance operator is separable, that is, if the following condition holds
\begin{align}
	\sigma(\z_{1},\z_{2}) & = \prod_{n=1}^{N} \sigma^{(n)}(z_{1}^{(n)},z_{2}^{(n)}), \quad \forall \z_{1},\z_{2} \in \domR^{N} \label{eq:separability_GRF}
\end{align}
where $\sigma^{(n)}: \domR \times \domR \mapsto \domR$ is the covariance operator specific to the $n$-th coordinate axis, and independent to the covariance operator on the other axes.

\begin{remark} \label{ex:bond}
	For contextual clarity, consider the random variable, $x(m,c) \in \domR$, which represents the fixed income return as a function of the maturity $m$ and country $c$. This can be viewed as a scalar field on a 2-dimensional coordinate system (\textit{maturity} $\times$ \textit{country}). The separability condition asserts that
	\begin{align}
		\cov{ x(m_{i},c_{j}), x(m_{l},c_{k}) } & = \sigma^{(m)}_{il} \sigma^{(c)}_{jk}
	\end{align}
	where $\sigma^{(m)}_{il}$ is the return covariance between the $i$-th and $l$-th maturities, and is independent of the countries. Similarly, $\sigma^{(c)}_{jk}$ is the return covariance between the $j$-th and $k$-th countries, and is independent of the maturities.
\end{remark}

\begin{figure}[h!]
	\centering
	\includegraphics[width=0.45\textwidth, trim={0cm 0 0 0},clip]{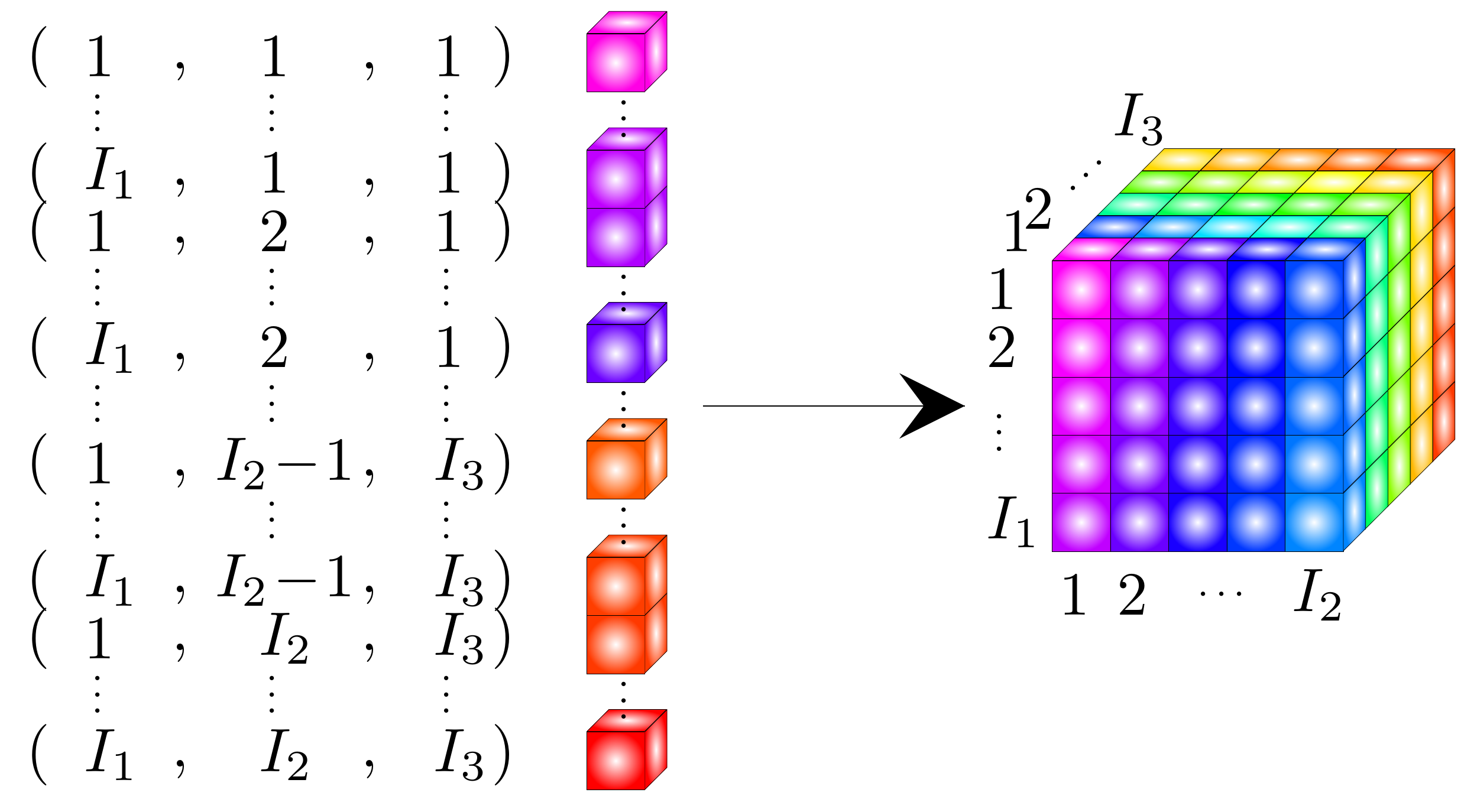}
	\caption{\label{fig:multi_index} Tensorization of scalar variables on a 3D coordinate system to form an order-3 tensor. Each variable possesses a coordinate $3$-tuple.}
\end{figure}

The act of forming the tensor $\calX \in \domR^{I_{1}\times \cdots \times I_{N}}$ from $K$ scalar-valued variables is known as \textit{tensorization}. The scalar-valued samples are ordered as follows
\begin{align}
[\calX]_{i_{1}...i_{N}} = x(z_{i_{1}}^{(1)},...,z_{i_{N}}^{(N)}), \!\! \quad i_{n} = 1,...,I_{n}, \!\! \quad z_{i_{n}}^{(n)} \in \domR
\end{align}
Figure \ref{fig:multi_index} illustrates the tensorization of scalar-valued random variables to form an order-3 tensor.

\vspace{-0.5cm}


Owing to the condition in (\ref{eq:separability_GRF}), the tensor $\calX$ is said to exhibit a \textit{Kronecker separable} covariance structure, and therefore has the following statistical properties \cite{Hoff2011,Scalzo2019_1}
\begin{align}
\expect{\|\calX\|^{2}} & = \sigma^{2} \label{eq:tensor_variance} \\
\expect{\X_{(n)}\X_{(n)}^{\Trans}} & = \sigma^{2}\boldTheta^{(n)} \label{eq:mode_n_covariance} \\
\expect{\x\x^{\Trans}} & = \sigma^{2}\left( \kronprod{n=N}{1} \boldTheta^{(n)} \right) \label{eq:kron_covariance}
\end{align}
where $\x = \vect{\calX}$ and $\|\cdot \|$ denotes the Frobenius norm. Intuitively, $\sigma^{2}$ is the average variance over all scalar-valued variables within the tensor, and $\boldTheta^{(n)} \in \domR^{I_{n} \times I_{n}}$ is the mode-$n$ \textit{covariance density} matrix, which has the unit-trace property $\tr{\boldTheta^{(n)}} = 1$, $\forall n$. The covariance density, $\boldTheta^{(n)}$, designates the proportion of the total variance, $\sigma^{2}$, allocated to each fibre.

The defining feature of the \textit{Kronecker separability} condition is that the covariance matrix, $\boldSigma$, is characterised in terms of \textit{fiber-to-fiber} (multilinear) covariance parameters, in contrast to element-to-element (linear) covariance parameters, as implied by the multivariate normal distribution.

In addition, the separability condition provides a stable and parsimonious alternative to an unrestricted estimate of $\boldSigma$, the latter being unstable or even unavailable if the dimensions of the sample tensor are large compared to the number of samples.

\begin{remark}
	Consider the covariance matrix, $\boldSigma \in \domR^{K \times K}$, which constitutes of $\frac{1}{2}\left( K^{2} + K \right)$ distinct parameters. In turn, the Kronecker separable counterpart, $\sigma^{2}\left( \kronprod{n=N}{1} \boldTheta^{(n)} \right)$, reduces to $1 + \frac{1}{2}\sum_{n=1}^{N} \left(I_{n}^{2} + I_{n}\right) $ distinct parameters. Referring back to Remark \ref{ex:bond}, consider the case where we observe fixed income returns for $I_{m}=15$ maturities and $I_{c}=8$ countries, that is, for $K=120$ returns in total. Then, the multivariate covariance matrix will have $\frac{1}{2}\left( 120^{2} + 120 \right) = 7260$ distinct parameters, whereas the Kronecker separable counterpart reduces the model to $1 + \frac{1}{2}\left( 15^{2} + 15 + 8^{2} + 8 \right) = 157$. The practical utility of such a parameter reduction is evident.
\end{remark}

\subsection{Multilinear principal component analysis}

Consider the eigendecomposition of the mode-$n$ covariance density matrix, that is
\begin{align}
	\boldTheta^{(n)} = \U^{(n)}\boldLambda^{(n)}\U^{(n)\Trans}
\end{align}
where $\U^{(n)} \in \domR^{I_{n} \times I_{n}}$ is the mode-$n$ eigenvector matrix, and $\boldLambda^{(n)} \in \domR^{I_{n} \times I_{n}}$ is the mode-$n$ eigenvalue matrix, which is diagonal and has unit-trace property, $\tr{\boldLambda^{(n)}}=1$, $\forall n$. Following from the properties of the Kronecker product \cite{Magnus1985}, we can express the Kronecker separability conditions in (\ref{eq:mode_n_covariance})-(\ref{eq:kron_covariance}) as follows
\begin{align}
	\expect{\X_{(n)}\X_{(n)}^{\Trans}} & = \sigma^{2}\, \U^{(n)} \boldLambda^{(n)} \U^{(n)\Trans}  \\
	\expect{\x\x^{\Trans}} & = \sigma^{2}\left( \kronprod{n=N}{1} \U^{(n)} \boldLambda^{(n)} \U^{(n)\Trans} \right) 
\end{align}
We refer to this result as the \textit{multilinear PCA} of the tensor-valued random variable $\calX$. This result is intrinsically linked to the well-known multilinear singular value decomposition, also referred to as the Tucker decomposition \cite{Tucker1966,DeLathauwer2000}.


\newpage

\section{Global Fixed Income Factor Analysis}

We next proceed to develop the multilinear model for global fixed income returns. The data structures considered in the sequel are order-$2$ tensors, that is, matrix-valued random variables, denotes by $\X \in \domR^{I_{1} \times I_{2}}$. The mode-$n$ unfoldings of order-$2$ tensors reduce to 
\begin{align}
	\X_{(1)} &= \X \quad \in \domR^{I_{1} \times I_{2}} \\
	\X_{(2)}& = \X^{\Trans} \quad \!\!\! \in \domR^{I_{2} \times I_{1}}
\end{align}
Although we focus our attention to order-2 tensors, it is important to note that the multilinear algebra tools provided in the previous section, and the model we develop next, naturally generalise to tensors of any order $N$.


\subsection{Data preparation}

Consider the \textit{i.i.d.} random variable, $x_{t}(m,c) \in \domR$, which represents the return of a fixed income asset with maturity $m$ and from the country $c$ at a time instant $t$. For simplicity, we can assume that the return is distributed according to
\begin{align}
	x_{t}(m,c) \sim \Normal{ 0, \sigma^{2}(m,c) }
\end{align}
which is dependent on the maturity and country.

When jointly considering the returns of $I_{m}$ maturities and $I_{c}$ countries, we can tensorize the collection of returns at each time instant $t$ to form the order-$2$ tensor-valued random variable $\X_{t} \in \domR^{I_{m} \times I_{c}}$, given by
\begin{align}
	[\X_{t}]_{ij} = x_{t}(m_{i},c_{j}), \quad i=1,..,I_{m}, \quad j=1,...,I_{c}, \label{eq:tensorize}
\end{align}
Each tensor has $I_{m}I_{c}$ returns points in total.

For clarity, it is important to understand the physical meaning of the vector and matrix representations of the considered tensor. First, define the $i$-th maturity fibre, $\f_{i}^{(m)} \in \domR^{I_{m}}$, and $i$-th country fibre, $\f_{i}^{(c)} \in \domR^{I_{c}}$, respectively as follows
\begin{align}
	\f_{i}^{(m)} = \left[ \begin{array}{c}
	x(m_{1},c_{i}) \\
	x(m_{2},c_{i}) \\
	\vdots \\
	x(m_{I_{m}},c_{i})
	\end{array} \right], \quad \f_{i}^{(c)} = \left[ \begin{array}{c}
	x(m_{i},c_{1}) \\
	x(m_{i},c_{2}) \\
	\vdots \\
	x(m_{i},c_{I_{c}})
	\end{array} \right]
\end{align}
In other words, the $i$-th maturity fibre, $\f_{i}^{(m)}$, contains the return of all maturities associated to the $i$-th country (the returns of an entire domestic curve). In turn, the $i$-th country fibre, $\f_{i}^{(c)}$, contains the returns of all countries associated to the $i$-th maturity.

\begin{figure}[h]
	\centering
	\includegraphics[width=0.5\textwidth, trim={0cm 0 0 0},clip]{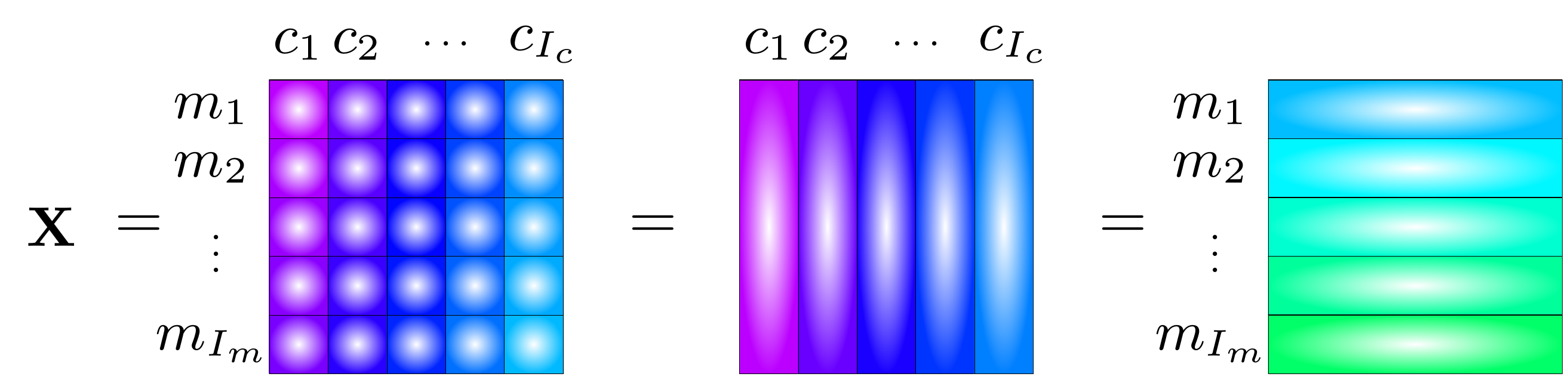}
	\caption{\label{fig:tensor_representations} Illustration of a tensor-valued sample, represented in terms of individual returns (left panel), maturity fibres, $\f_{i}^{(m)}$, (middle panel) and country fibres, $\f_{i}^{(c)}$, (right panel).}
\end{figure}

With that, the tensor can be expressed in terms of the maturity fibres, or equivalently through the country fibres, as follows
\begin{align}
\X  = \left[ \begin{array}{cccc}
\f_{1}^{(m)} & \f_{2}^{(m)} & \cdots & \f_{I_{c}}^{(m)}
\end{array} \right] = \left[ \begin{array}{c}
\f_{1}^{(c) \Trans} \\
\f_{2}^{(c) \Trans} \\
\vdots \\
\f_{I_{m}}^{(c) \Trans} \\
\end{array} \right] \label{eq:tensor_representation_fibres}
\end{align}
Refer to Figure \ref{fig:tensor_representations} for an illustrative description.


\pagebreak

Similarly, the vector representation can be visualised, and written, in terms of maturity fibres (domestic curves), that is \vspace{-0.2cm}

\begin{minipage}{0.23\textwidth}
	\vspace{0.25cm}
	\begin{flushright}
		\includegraphics[width=0.4\textwidth, trim={0cm 0 0 0},clip]{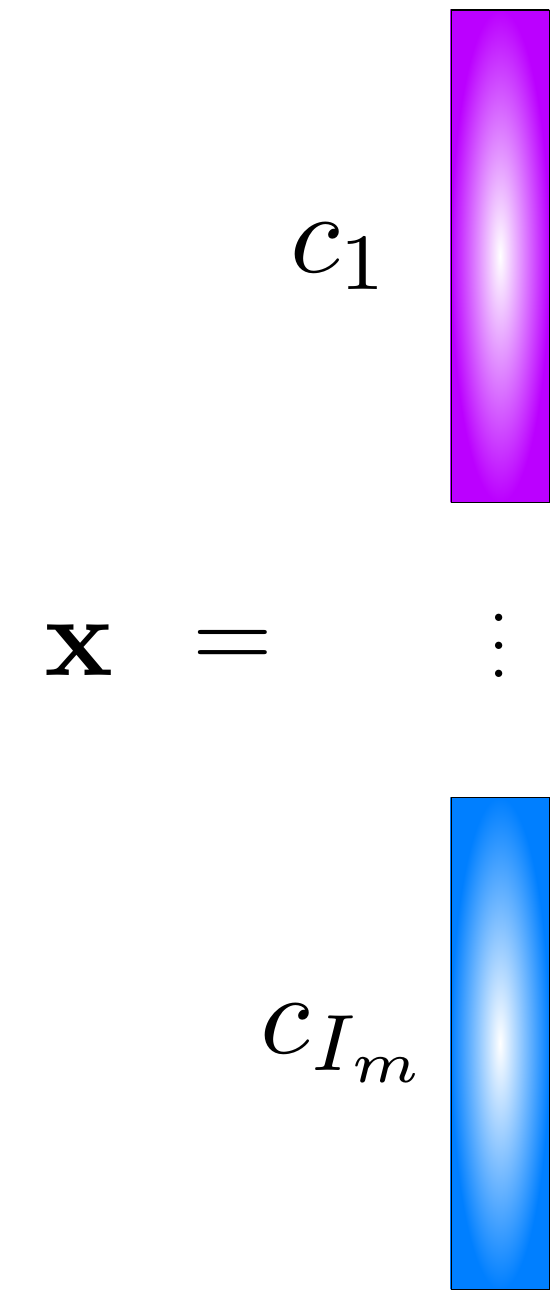}
	\end{flushright}
\end{minipage}
\begin{minipage}{0.23\textwidth}
	\begin{flalign}
	\hspace{-1.75cm}
	= \,\,\, \left[\begin{array}{c}
	\f_{1}^{(m)} \\ 
	\f_{2}^{(m)} \\
	\vdots \\
	\f_{I_{c}}^{(m)}
	\end{array}\right]
	\end{flalign}
\end{minipage}


\vspace{-0.5cm}

\subsection{Kronecker separability assumptions}

We next show that the multilinear algebra allows us to naturally decompose the multivariate covariance matrix into two parallel covariance matrices -- \textit{maturity-domain} and \textit{country-domain} covariance. This is possible owing to the defining  Kronecker separability feature of tensor-valued models in (\ref{eq:tensor_variance})-(\ref{eq:kron_covariance}), which in this case reduce to
\begin{align}
	\expect{\|\X\|^{2}} & = \sigma^{2} \\
	\expect{\X\X^{\Trans}} & = \sigma^{2}\boldTheta^{(m)} \\
	\expect{\X^{\Trans}\X} & = \sigma^{2}\boldTheta^{(c)} \\
	\boldSigma = \expect{\x\x^{\Trans}} & = \sigma^{2} \left(\boldTheta^{(c)} \! \otimes \! \boldTheta^{(m)} \right) \label{eq:kron_global}
\end{align}
Intuitively, $\sigma^{2}$ is the average variance of all fixed income returns, $\boldTheta^{(m)} \in \domR^{I_{m} \times I_{m}}$ is the \textit{maturity-domain} covariance density matrix, and $\boldTheta^{(c)} \in \domR^{I_{c} \times I_{c}}$ is \textit{country-domain} covariance density matrix. 

Using the tensor representation in (\ref{eq:tensor_representation_fibres}) based on fibres, it is clear that $\boldTheta^{(m)}$ and $\boldTheta^{(c)}$ respectively describe the average maturity-to-maturity and country-to-country covariance, since we can employ the total expectation theorem \cite{Weiss2005} to show that
\begin{align}
	\expect{\X\X^{\Trans}} & = E_{i}\!\left\{ \expect{ \f_{i}^{(m)}\f_{i}^{(m)\Trans} } \right\} \\
	\expect{\X^{\Trans}\X} & = E_{i}\!\left\{ \expect{ \f_{i}^{(c)}\f_{i}^{(c)\Trans} } \right\}
\end{align}
where $E_{i}\{\cdot\}$ denotes the expectation over the indices $i$. Equivalently, we can inspect the elements of the covariance matrices
\begin{align}
	[\expect{\X\X^{\Trans}}]_{kl} & = E_{i}\!\left\{ \cov{x(m_{k},c_{i}),x(m_{l},c_{i})} \right\} \\
	[\expect{\X^{\Trans}\X}]_{kl} & = E_{i}\!\left\{ \cov{x(m_{i},c_{k}),x(m_{i},c_{l})} \right\}
\end{align}
Therefore the statistic $[\expect{\X\X^{\Trans}}]_{kl}$ describes the \textit{expected} return covariance between all assets with maturity $m_{k}$ and $m_{l}$. Similarly, $[\expect{\X^{\Trans}\X}]_{kl}$ describes the \textit{expected} return covariance between all assets within countries $c_{k}$ and $c_{l}$.

\subsection{Implied domestic and cross-country dynamics}

Equipped with the Kronecker separability conditions, we next investigate the domestic and cross-country fixed income return interactions implied by the multilinear model.

We begin by observing that the multivariate covariance matrix is in fact a block matrix of domestic and cross-country fixed income return covariance matrices, that is
\begin{align}
\boldSigma = \left[ \begin{array}{cccc}
\boldSigma_{11} & \boldSigma_{12} & \cdots & \boldSigma_{1I_{c}} \\
\boldSigma_{21} & \boldSigma_{22} & \ddots & \vdots \\
\vdots & \ddots & \ddots & \vdots \\
\boldSigma_{I_{c}1} & \cdots & \cdots & \boldSigma_{I_{c}I_{c}}
\end{array} \right]
\end{align}
where $\boldSigma_{ii} \in \domR^{I_{m} \times I_{m}}$ is the \textit{domestic} covariance matrix of the $i$-th country, and $\boldSigma_{ij} \in \domR^{I_{m} \times I_{m}}$ is the cross-country covariance matrix between countries $i$ and $j$. 

From the Kronecker separability condition in (\ref{eq:kron_global}), we can also express the multivariate covariance matrix as the following block matrix
\begin{align}
\boldSigma & = \left[ \begin{array}{cccc}
\sigma^{2}\theta_{11}^{(c)}\boldTheta^{(m)} & \sigma^{2}\theta_{12}^{(c)}\boldTheta^{(m)} & \cdots & \sigma^{2}\theta_{1I_{c}}^{(c)}\boldTheta^{(m)} \\
\sigma^{2}\theta_{21}^{(c)}\boldTheta^{(m)} & \sigma^{2}\theta_{22}^{(c)}\boldTheta^{(m)} & \ddots & \vdots \\
\vdots & \ddots & \ddots & \vdots \\
\sigma^{2}\theta_{I_{c}1}^{(c)}\boldTheta^{(m)} & \cdots & \cdots & \sigma^{2}\theta_{I_{n}I_{c}}^{(c)}\boldTheta^{(m)}
\end{array} \right]
\end{align}
where $\theta_{ij}^{(c)}$ is the $(i,j)$-th element of $\boldTheta^{(c)}$. Therefore the multilinear model asserts that each domestic and cross-country covariance matrix takes the form
\begin{align}
\boldSigma_{ij} = \sigma^{2}\theta_{ij}^{(c)}\boldTheta^{(m)} \label{eq:cross-country_covariance}
\end{align}
The multilinear model implicitly assumes that the maturity-domain covariance, given by $\boldTheta^{(m)}$, is identical for all countries. This is to say that all countries exhibit the same \textit{level}, \textit{slope} and \textit{curvature} factors, as is demonstrated empirically in the Section \ref{section:analysis}. Moreover, the term $\theta_{ij}^{(c)}$ simply scales the variance parameter, $\sigma^{2}$, so as to match the observed cross-country variance, i.e. $\tr{\boldSigma_{ij}}=\sigma^{2}\theta_{ij}^{(c)}$.

\subsection{Multilinear factor analysis}

It is natural to next evaluate and interpret the orthogonal bases spanned by the maturity-domain and country-domain covariance density matrices, which are obtained through the following eigendecompositions
\begin{align}
\boldTheta^{(m)} & = \U^{(m)}\boldLambda^{(m)}\U^{(m)\Trans} \label{eq:EVD_maturity} \\
\boldTheta^{(c)} & = \U^{(c)}\boldLambda^{(c)}\U^{(c)\Trans} \label{eq:EVD_country}
\end{align}
The maturity-domain eigenvector matrix, $\U^{(m)} \in \domR^{I_{m}\times I_{m}}$, contains vectors $\u_{i}^{(m)} \in \domR^{I_{m}}$ which describe orthogonal directions in maturity-to-maturity covariance. These vectors represent the well-known level, slope and curvature factors. Similarly, $\U^{(c)} \in \domR^{I_{c}\times I_{c}}$ contains vectors $\u_{i}^{(c)} \in \domR^{I_{c}}$ which describe orthogonal directions in country-to-country covariance. These also have economic meaning, as is shown in the sequel. 

Note that the eigenvector matrices are orthogonal, that is, $\U^{(m)\Trans}\U^{(m)} = \I$ and $\U^{(c)\Trans}\U^{(c)} = \I$.

The associated eigenvalue matrices, $\boldLambda^{(m)} \in \domR^{I_{m}\times I_{m}}$ and $\boldLambda^{(c)} \in \domR^{I_{c}\times I_{c}}$, respectively hold the eigenvalues, $\lambda_{i}^{(m)}$ and $\lambda_{i}^{(c)}$, which describe the fraction of the total variance, $\sigma^{2}$, explained by factor $\u_{i}^{(m)}$ and $\u_{i}^{(c)}$. As such, the eigenvalues sum up to unity, $\tr{\boldLambda^{(m)}}=\tr{\boldLambda^{(c)}} = 1$.

Furthermore, we can describe the eigenvectors and eigenvalues of the multivariate covariance matrix, $\boldSigma$, in terms of maturity-domain and country-domain spectral parameters. Upon decomposing the covariance matrix as
\begin{align}
	\boldSigma = \U\boldLambda\U^{\Trans}
\end{align}
we can show that the following relationships hold
\begin{align}
	\U & = \left( \U^{(c)} \! \otimes \! \U^{(m)} \right) \\
	\boldLambda & = \sigma^{2} \left( \boldLambda^{(c)} \! \otimes \! \boldLambda^{(m)} \right)
\end{align}
By inspecting each eigenvector $\u_{i} \in \domR^{I_{m}I_{c}}$ within $\U \in \domR^{I_{m}I_{c} \times I_{m}I_{c}}$, as well as its corresponding eigenvalue $\lambda_{i}$, we can obtain the following relationships \vspace{-0cm}
\begin{align}
	\u_{i} & = \left( \u_{k}^{(c)} \otimes \u_{l}^{(m)} \right) = \left[ \begin{array}{c}
	u_{k1}^{(c)} \, \u_{l}^{(m)} \\
	u_{k2}^{(c)} \, \u_{l}^{(m)} \\
	\vdots \\
	u_{kI_{c}}^{(c)} \, \u_{l}^{(m)} 
	\end{array} \right] \label{eq:eigenvector} \\
	\lambda_{i} & = \lambda_{k}^{(c)} \lambda_{l}^{(m)}
\end{align}
where $u_{kj}^{(c)}$ is the $j$-th element in the vector $\u_{k}^{(c)}$, and owing to the Kronecker properties, $i=(k-1)I_{m}+l$.

\begin{remark}
	Notice the repeated pattern in (\ref{eq:eigenvector}). The Kronecker separable structure asserts that, for a given factor $\u_{i}$, the maturity-domain structure within each country is the same, and equal to $\u_{l}^{(m)}$. This follows from the result in (\ref{eq:cross-country_covariance}). Another, less obvious, property is that the country-domain structure is the same across any maturity, and equal to $\u_{k}^{(c)}$.
\end{remark}

\vspace{-0.5cm}

\subsection{Estimation procedure}

Unlike existing approaches for joint term structure modeling, the proposed framework is entirely analytical, and so is the estimation procedure. The estimate of the variance parameter, $\sigma^{2}$, over $T$ time instants is given by
\begin{align}
	\sigma^{2} = \frac{1}{T-1}\sum_{t=1}^{T} \|\X_{t}\|^{2} \label{eq:estimate_variance}
\end{align}
Similarly, the maturity-domain and country-domain covariance density matrices are obtained as follows
\begin{align}
	\boldTheta^{(m)} & = \frac{1}{\sigma^{2}(T-1)}\sum_{t=1}^{T} \X_{t}\X_{t}^{\Trans} \\
	\boldTheta^{(c)} & = \frac{1}{\sigma^{2}(T-1)}\sum_{t=1}^{T} \X_{t}^{\Trans}\X_{t} \label{eq:estimate_country}
\end{align}
These are the maximum likelihood estimators of the tensor-valued Gaussian distribution, which have been shown to be statistically consistent \cite{Scalzo2019_1}.

\newpage

\section{Global Portfolio Management and Hedging}

The direct application of domestic principal components for evaluating market risk and constructing hedged portfolios has been widely studied and implemented in the financial industry \cite{Solomon2000,CreditSuisse2012,StandardChartered2013,TDSecurities2015}. However, understanding the commonalities between different country's term structures is also important for assessing the potential for international diversification and managing the risk of global fixed income portfolios. 

The portfolio risk measure we consider is the \textit{portfolio variance}, which is a function of the multivariate covariance matrix, $\boldSigma \in \domR^{I_{m}I_{c} \times I_{m}I_{c}}$. Given a vector of portfolio weights, $\w \in \domR^{I_{m}I_{c}}$, the portfolio variance is given by
\begin{align}
	\sigma_{p}^{2} = \w^{\Trans}\boldSigma\w
\end{align}
We have demonstrated in the previous section that, when considering an international basket of fixed income assets, the covariance, $\boldSigma$, exhibits the Kronecker separable structure in (\ref{eq:kron_global}). In light of this, we can show that if we choose our portfolio vector to match the Kronecker separable structure, that is, if we set
\begin{align}
	\w = \left( \w^{(c)} \otimes \w^{(m)} \right)
\end{align}
where $\w^{(m)} \in \domR^{I_{m}}$ and $\w^{(c)} \in \domR^{I_{c}}$ are respectively the maturity-domain and country-domain weights, then we arrive at parsimonious and compact solutions for global portfolio management. This owes to the reduction in parameters required for portfolio optimization from $I_{m}I_{c}$ to $(I_{m}+I_{c})$. This evident advantage is only achieved via the decomposition of overall risk into parallel domain -- maturity and country.


\subsection{Minimum variance portfolio}

The capital-constrained portfolio which attains the minimum variance is obtained through the optimization problem
\begin{align}
	\min_{\w} \,\, \sigma_{p}^{2}, \quad \textrm{s.t.} \,\, \w^{\Trans}\,\1 = 1
\end{align}
the solution of which is given by the well known minimum variance portfolio
\begin{align}
	\w = \frac{\boldSigma^{-1}\1}{\1^{\Trans}\boldSigma^{-1}\1}
\end{align}
Notice that for the Kronecker separable case the optimal portfolio reduces to
\begin{align}
	\w & = \frac{\left(\boldTheta^{(c)} \otimes \boldTheta^{(m)} \right)^{-1}\1}{\1^{\Trans}\left(\boldTheta^{(c)} \otimes \boldTheta^{(m)} \right)^{-1}\1} \notag\\
	& = \left( \frac{\boldTheta^{(c)-1}\1}{\1^{\Trans}\boldTheta^{(c)-1}\1} \otimes \frac{\boldTheta^{(m)-1}\1}{\1^{\Trans}\boldTheta^{(m)-1}\1} \right)
\end{align}
This results asserts that the portfolio optimization can be separated into parallel problems within the maturity and country domains. This is equivalent to solving for $\w^{(m)}$ and $\w^{(c)}$ independently through the following minimizations
\begin{align}
	\min_{\w^{(m)}} \,\, \w^{(m)\Trans}\boldTheta^{(m)}\w^{(m)}, \quad  \textrm{s.t.} \,\, \w^{(m)\Trans}\,\1 = 1 \notag
\end{align}
and
\begin{align}
	\min_{\w^{(c)}} \,\, \w^{(c)\Trans}\boldTheta^{(c)}\w^{(c)}, \quad \textrm{s.t.} \,\, \w^{(c)\Trans}\,\1 = 1 \notag
\end{align}

\subsection{Hedging}

Hedging of fixed income securities remains one of the most challenging problems faced by financial institutions. The sensitivity of a portfolio, $\w$, with respect to a risk factor, $\u$, is simply given by the inner product $\u^{\Trans}\w$. The aim is to form a portfolio which is orthogonal to the risk factor, that is, the aim is to attain $\u^{\Trans}\w=0$.

Within the considered international setup, risk factors are Kronecker separable, that is, $\u = \left( \u^{(c)}\otimes \u^{(m)} \right)$. Upon setting the portfolio vector to match the Kronecker separable structure, the risk exposure simplifies to
\begin{align}
	\u^{\Trans}\w & = \left( \u^{(c)\Trans}\w^{(c)} \right)\left( \u^{(m)\Trans}\w^{(m)} \right)
\end{align}
This result asserts that orthogonality, $\u^{\Trans}\w=0$, can be attained independently in either the maturity or country domain, and thus it is not necessary to do so in both simultaneously. In other words, orthogonality is achieved by either attaining $\u^{(c)\Trans}\w^{(c)}=0$ or $\u^{(m)\Trans}\w^{(m)}=0$. We next consider real-world applications of this result.


\subsubsection{Hedging long-term bonds}

Consider hedging a long-only portfolio of international long-term fixed income assets with $i$-th maturity (e.g. 30 years), using an international portfolio of shorter-term assets. The hedged portfolio must satisfy the following constraints within the maturity domain only:
\begin{align}
	\bolddelta_{i}^{\Trans} \w^{(m)} & = 1 \\
	\1^{\Trans}\w^{(m)} & = 0 \\
	\U^{(m)\Trans} \w^{(m)} & = \0
\end{align}
where $\bolddelta_{i} \in \domR^{I_{m}}$ is a vector of zeros with the $i$-th element equal to 1. Intuitively, the first condition reflects the long-only position in the asset with the $i$-th maturity, the second constrains the strategy to be self-financing, while the last enforces orthogonality with the maturity-domain factors.

\subsubsection{Hedging domestic bonds}

Conversely, consider hedging a domestic long-only portfolio within the $i$-th country, using an international portfolio. The hedged portfolio must satisfy the following constraints within the country domain only: \vspace{-0.1cm}
\begin{align}
	\bolddelta_{i}^{\Trans} \w^{(c)} & = 1 \\
	\1^{\Trans}\w^{(c)} & = 0 \\
	\U^{(c)\Trans} \w^{(c)} & = \0
\end{align} \vspace{-0.6cm} \\
where $\bolddelta_{i} \in \domR^{I_{c}}$. In this case, the first condition reflects the long-only position in the $i$-th country, the second constrains the strategy to be self-financing, while the last condition enforces orthogonality with the country-domain factors.

The portfolio hedging problems reduce to solving the linear systems $\A^{(m)} \w^{(m)} = \b^{(m)}$ and $\A^{(c)} \w^{(c)} = \b^{(c)}$, respectively. The optimal maturity-domain and country-domain weights are given by $\w^{(m)} = \A^{(m)+}\b^{(m)}$ and $\w^{(c)} = \A^{(c)+}\b^{(c)}$, where $(\cdot)^{+}$ denotes the Moore-Penrose inverse operator.

%
%
%
%
%
%
%
%
%
%

%
%
%

\newpage

\section{Empirical Analysis}

We next provide an empirical analysis of the global term structure of the international \textit{interest rate swaps} (IRS) market using the proposed multilinear factor model. The data comprised of weekly IRS rate curves\footnote{The \textit{swap rate} is the fixed interest rate that the receiver of the IRS demands in exchange for the uncertainty of having to pay the short-term floating LIBOR rate over time.}, ranging in the period 2015-01-01 to 2019-07-01, for eight developed economies, including Switzerland, Euro Area, United Kingdom, Japan, Australia, New Zealand, Canada and United States. Each domestic IRS curve consisted of swaps with maturities $\{1,2,3,4,5,6,7,8,9,10,12,15,20,25,30\}$ years. Therefore, at each time instant, we observed $I_{m}=15$ IRS returns for each of the $I_{c}=8$ economies, resulting in $I_{m}I_{c}=120$ daily observations in total. Figure \ref{fig:domestic_PCA} displays the historical IRS rates employed in the analysis. A glance at the collective behaviour of the historical data helps us understand the importance of global factors in driving the co-movement of fixed income securities across advanced economies.

\subsection{Domestic analysis}

\begin{figure}[H]
	\captionsetup[subfigure]{aboveskip=0pt,belowskip=5pt}
	
	\centering
	
	\begin{subfigure}[t]{0.5\textwidth}
		\centering
		\includegraphics[width=0.49\textwidth,trim={0 1.2cm 0 0},clip]{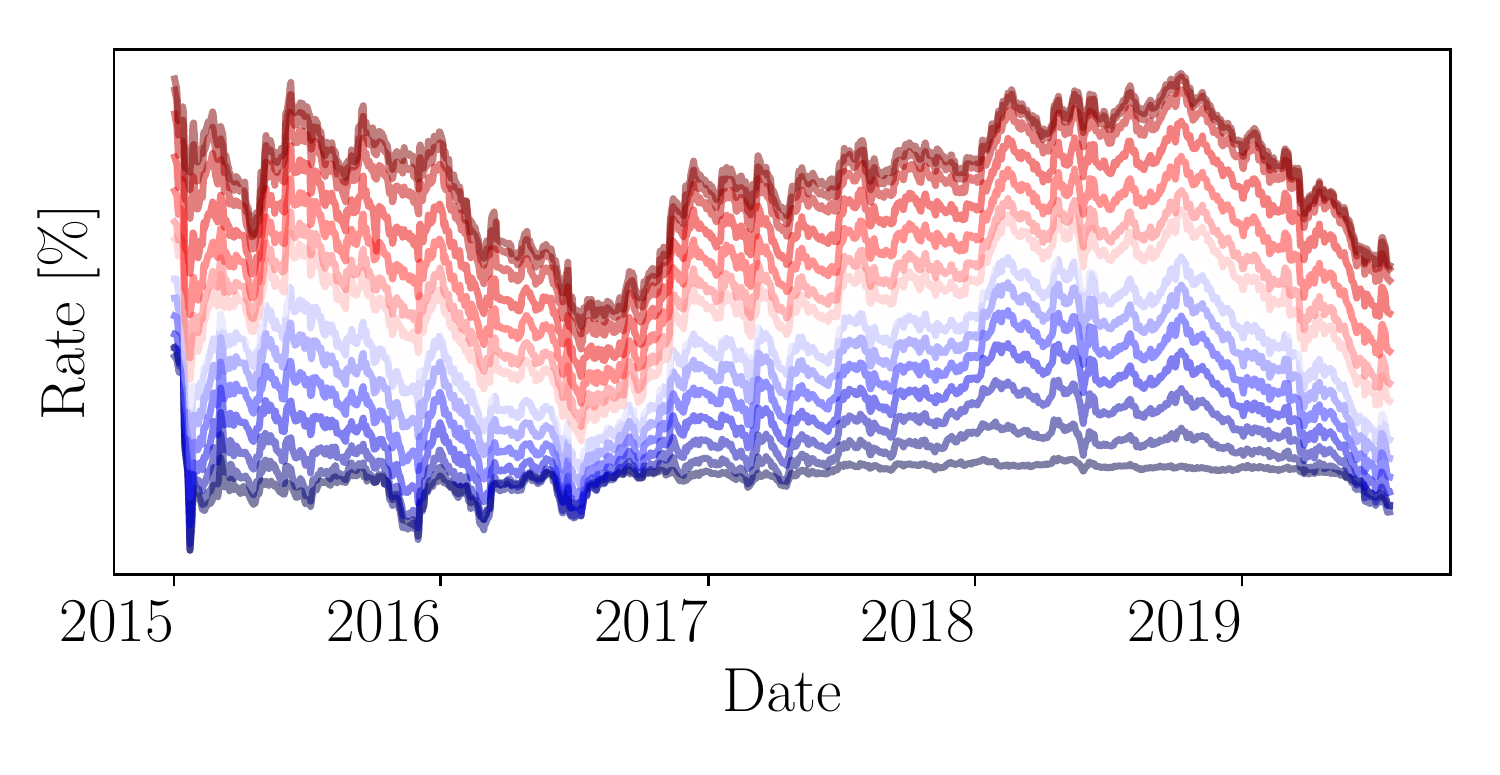}
		\includegraphics[width=0.49\textwidth,trim={0 1.2cm 0 0},clip]{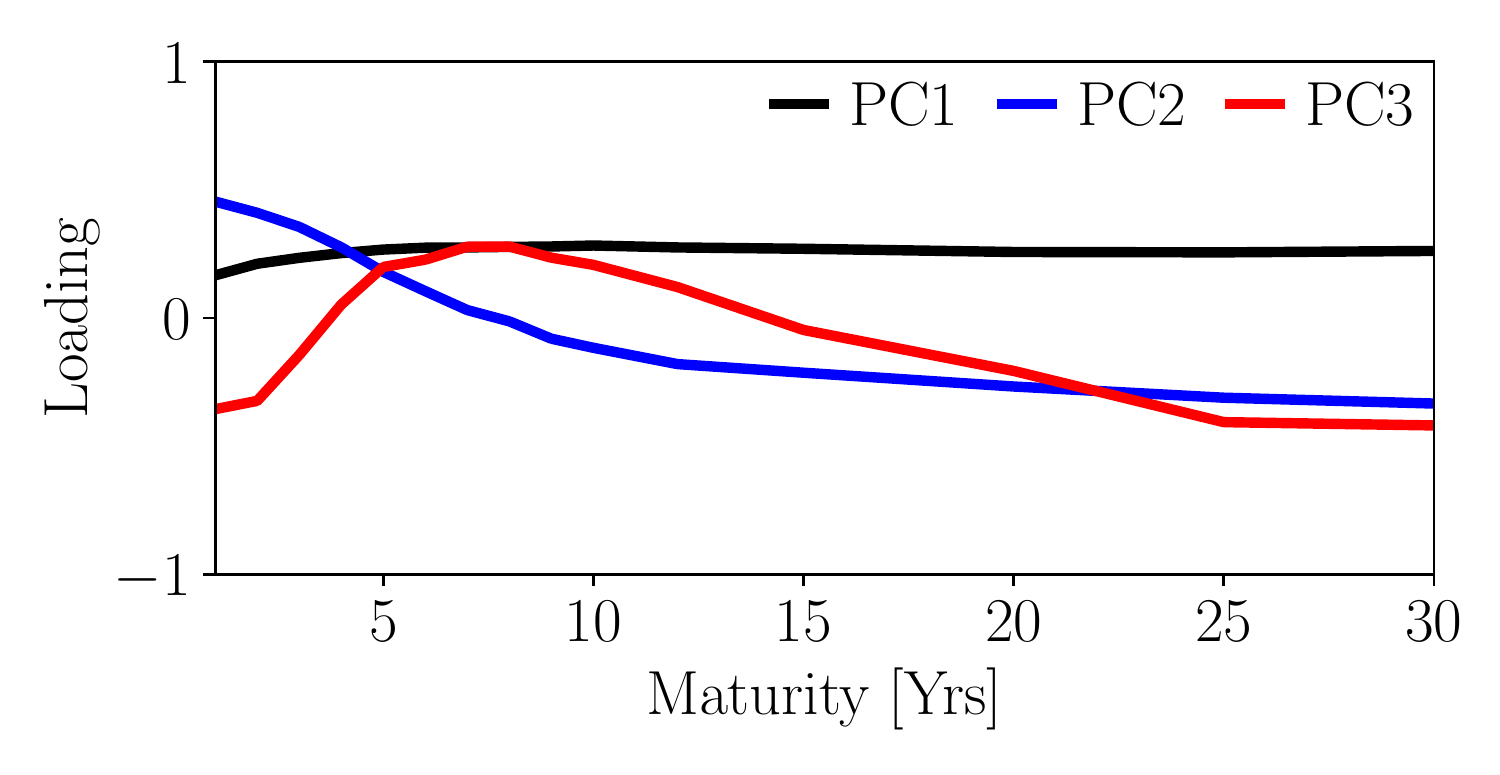}
		\caption{Switzerland (SF)}    
	\end{subfigure}
	
	\begin{subfigure}[t]{0.5\textwidth}
		\centering
		\includegraphics[width=0.49\textwidth,trim={0 1.2cm 0 0},clip]{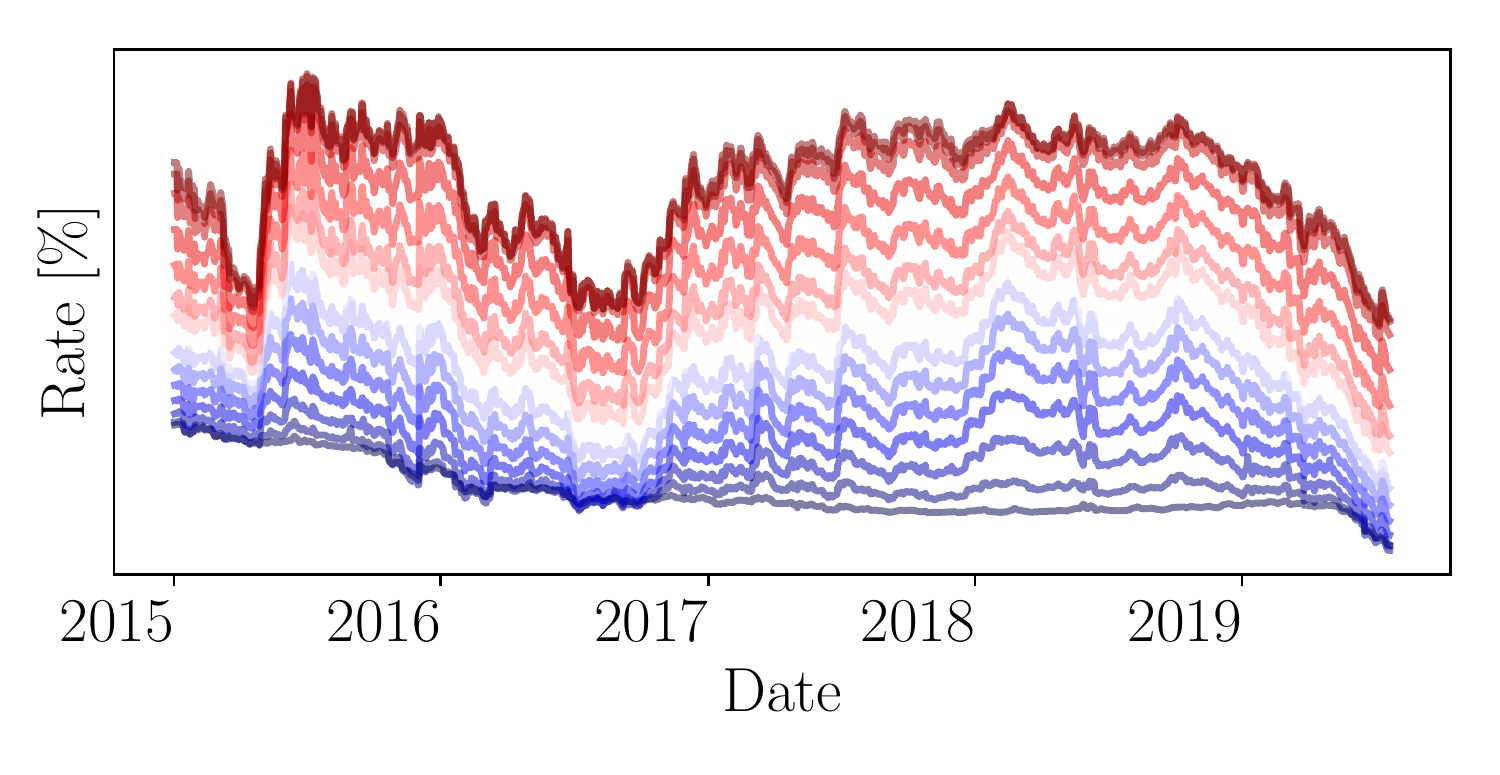}
		\includegraphics[width=0.49\textwidth,trim={0 1.2cm 0 0},clip]{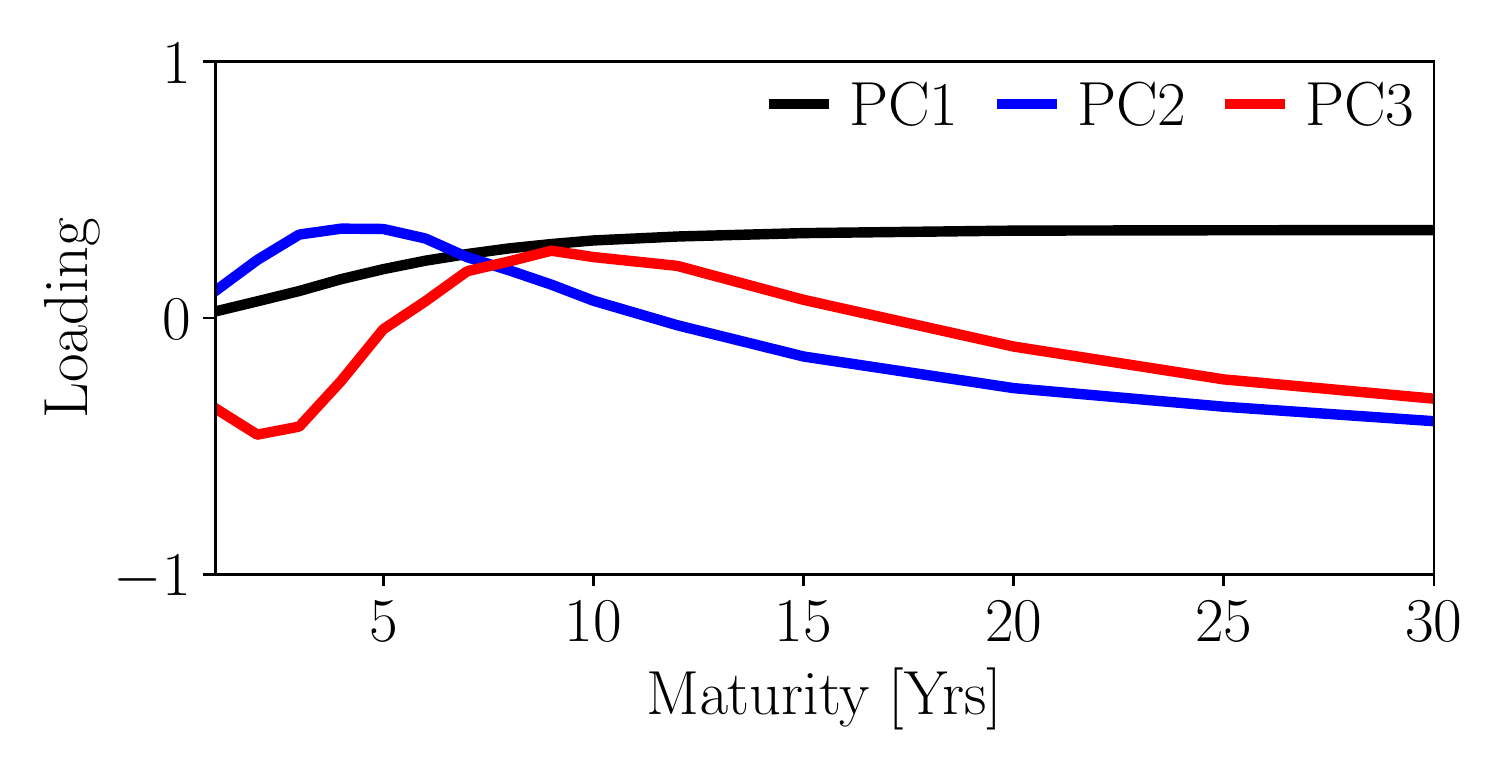}
		\caption{Euro Area (EU)}    
	\end{subfigure}

	\begin{subfigure}[t]{0.5\textwidth}
		\centering
		\includegraphics[width=0.49\textwidth,trim={0 1.2cm 0 0},clip]{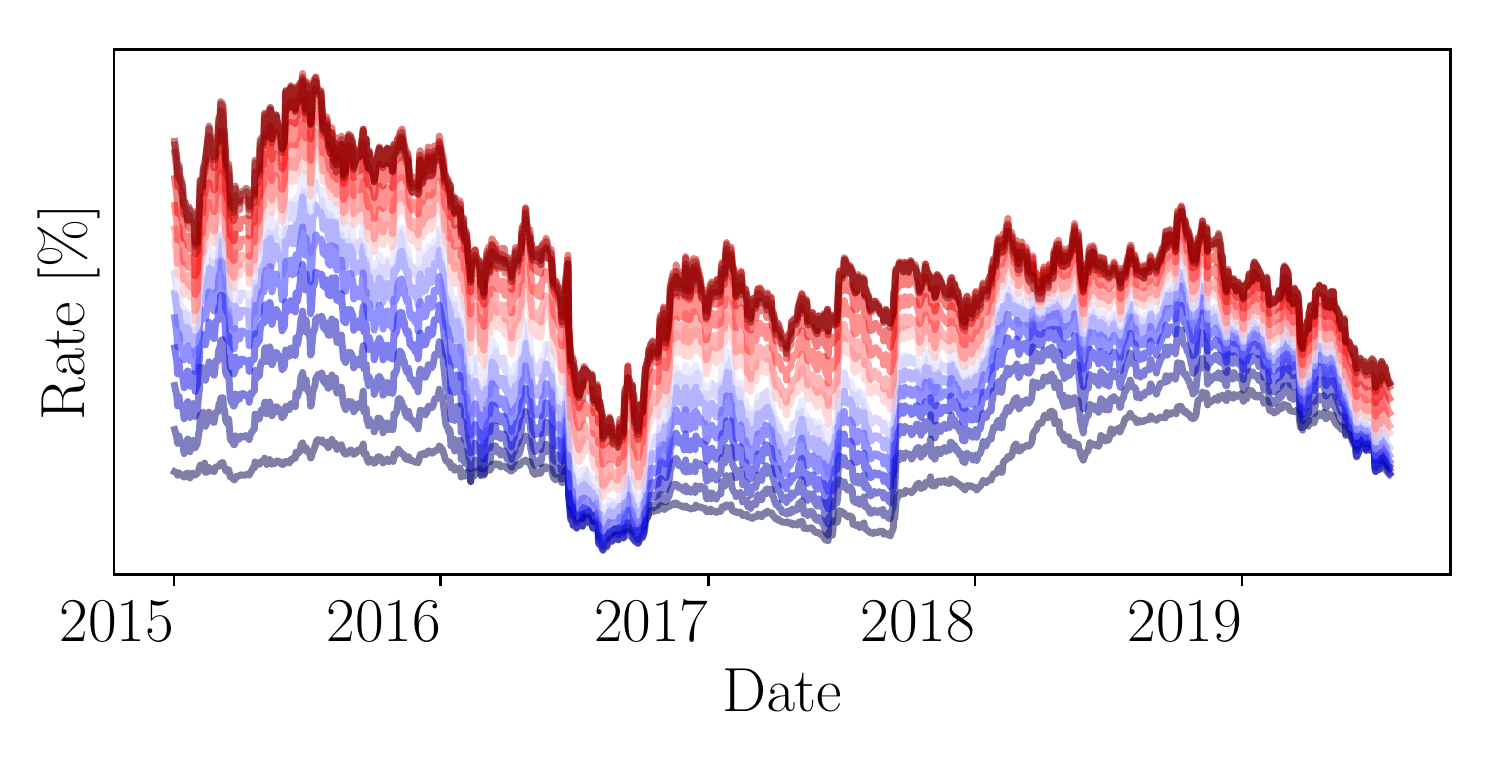}
		\includegraphics[width=0.49\textwidth,trim={0 1.2cm 0 0},clip]{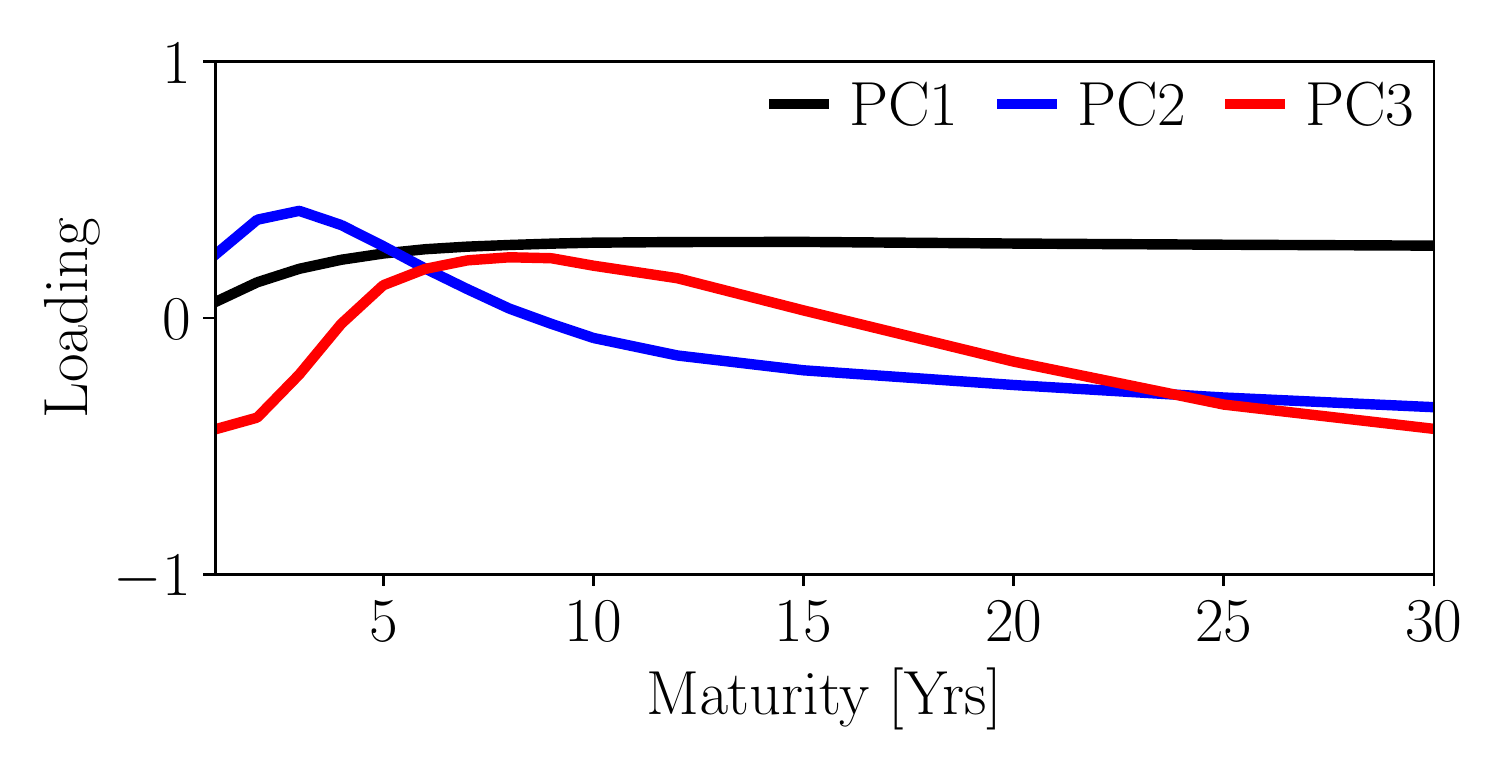}
		\caption{Great Britain (GB)}    
	\end{subfigure}
	
	\begin{subfigure}[t]{0.5\textwidth}
		\centering
		\includegraphics[width=0.49\textwidth,trim={0 1.2cm 0 0},clip]{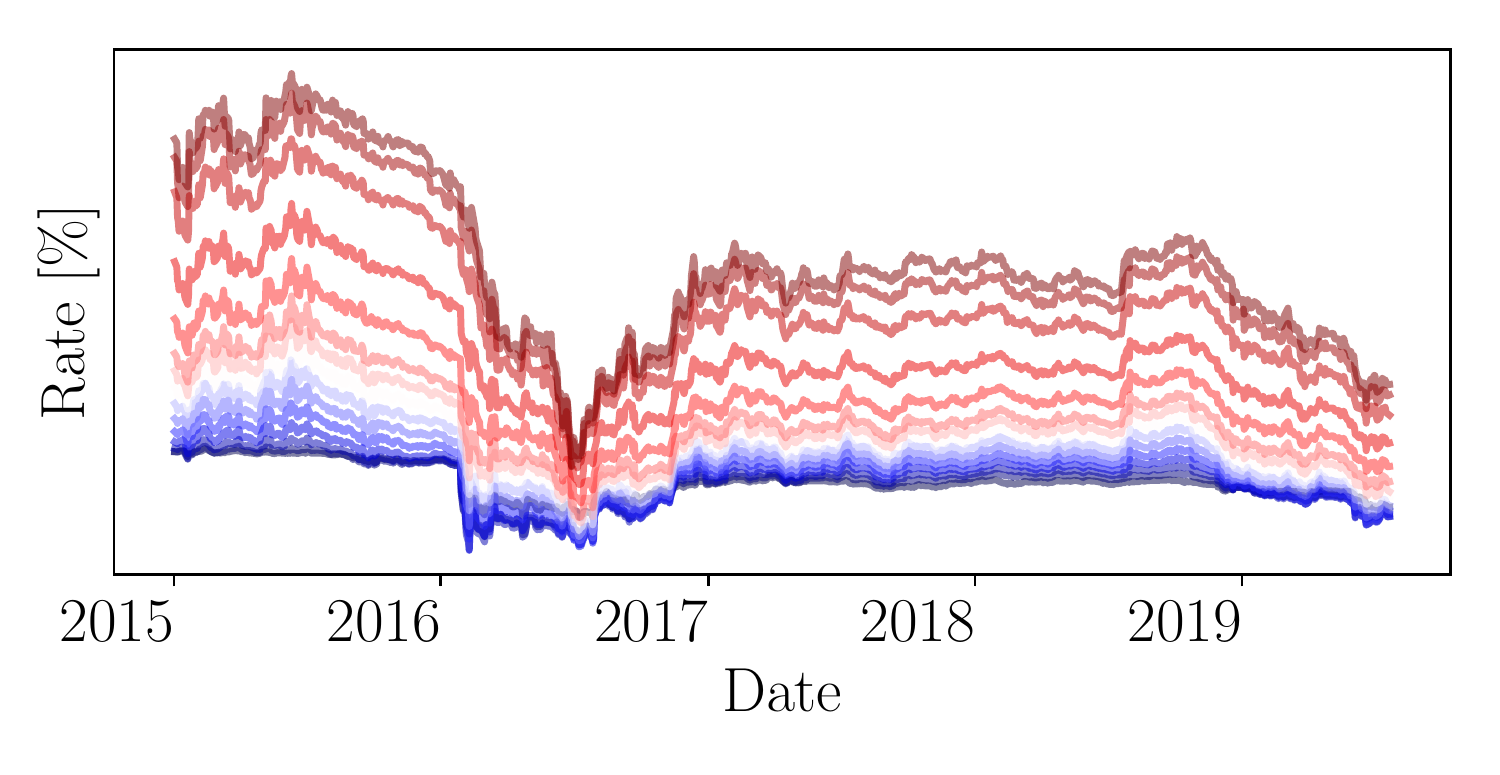}
		\includegraphics[width=0.49\textwidth,trim={0 1.2cm 0 0},clip]{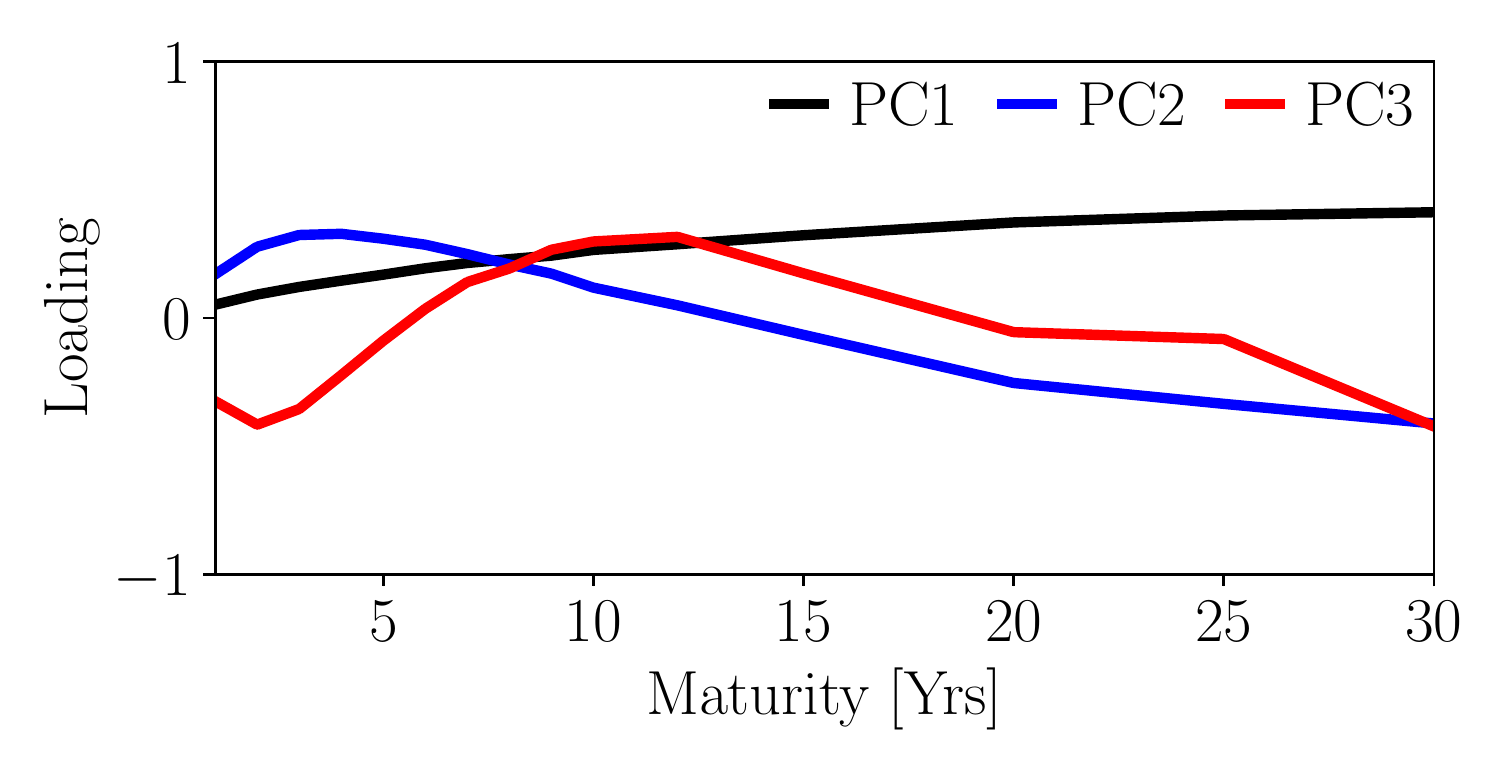}
		\caption{Japan (JP)}    
	\end{subfigure}
	
	\begin{subfigure}[t]{0.5\textwidth}
		\centering
		\includegraphics[width=0.49\textwidth,trim={0 1.2cm 0 0},clip]{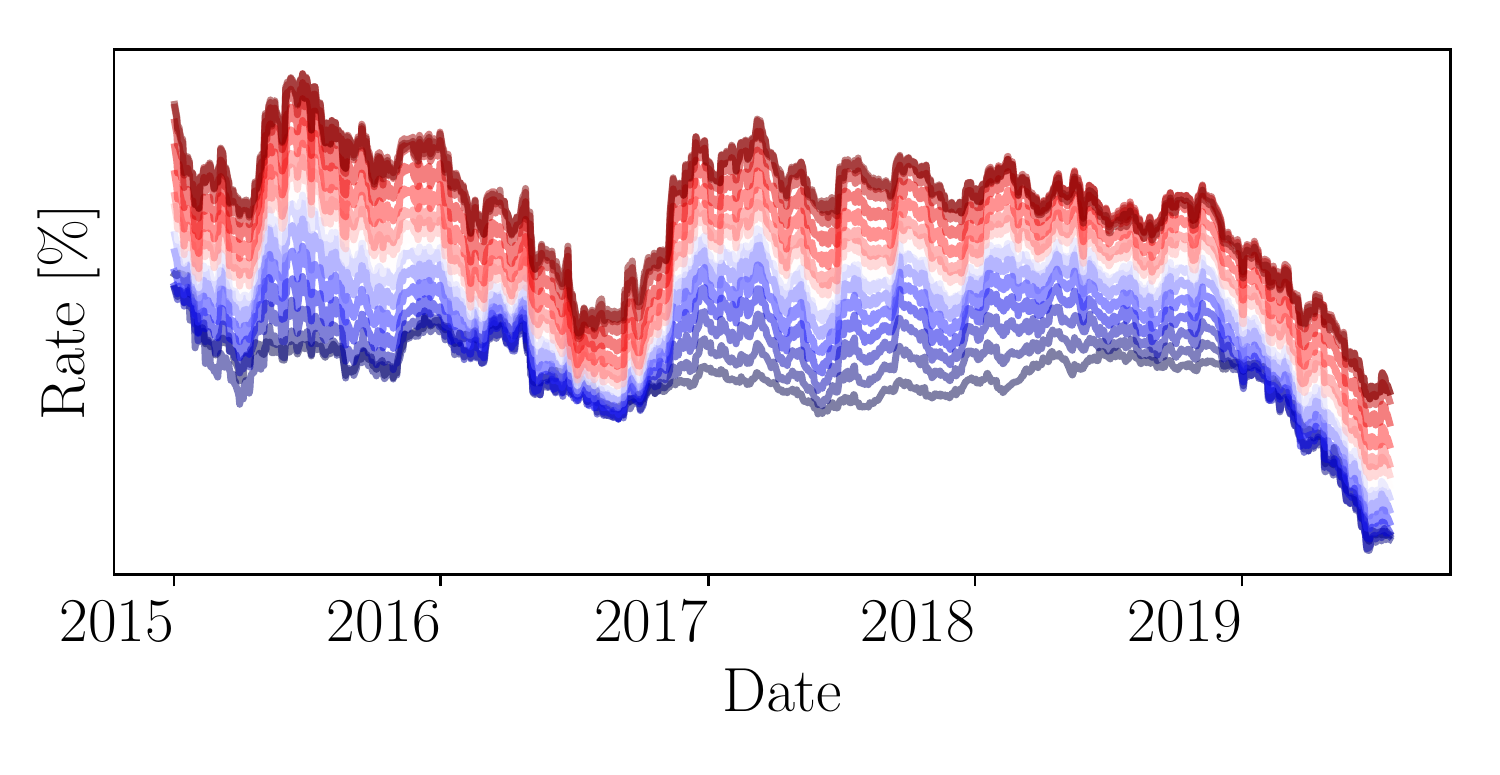}
		\includegraphics[width=0.49\textwidth,trim={0 1.2cm 0 0},clip]{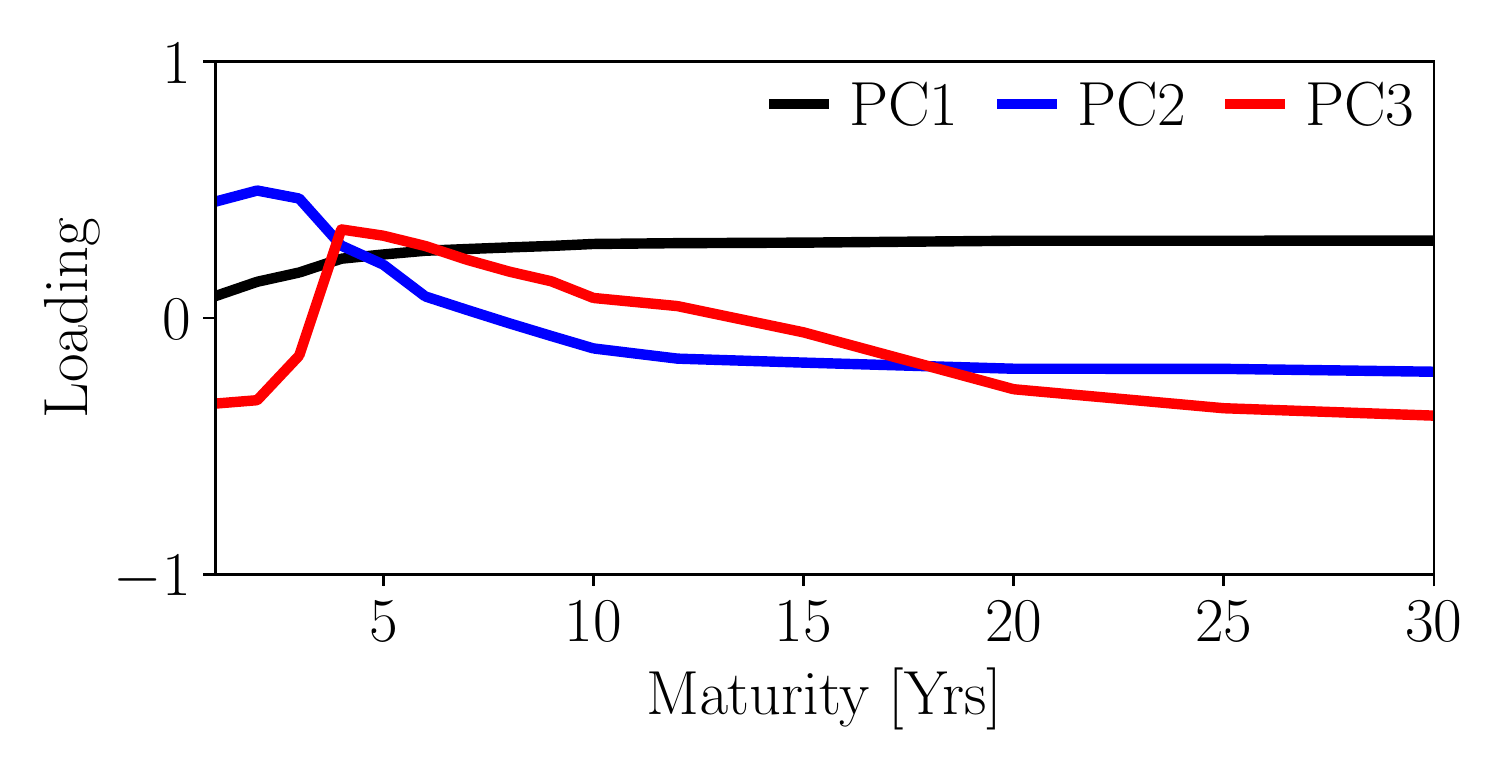}
		\caption{Australia (AU)}    
	\end{subfigure}
	
	\begin{subfigure}[t]{0.5\textwidth}
		\centering
		\includegraphics[width=0.49\textwidth,trim={0 1.2cm 0 0},clip]{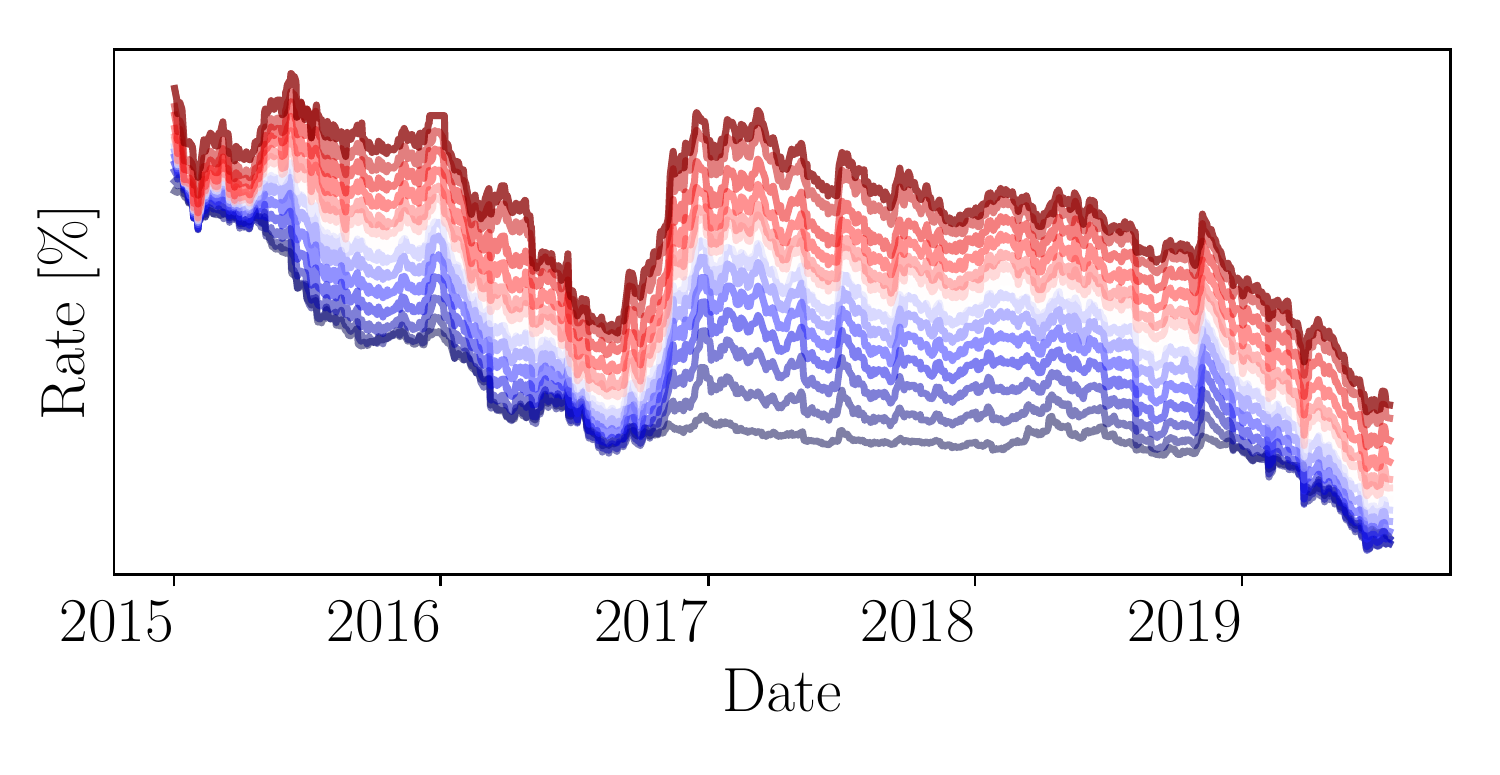}
		\includegraphics[width=0.49\textwidth,trim={0 1.2cm 0 0},clip]{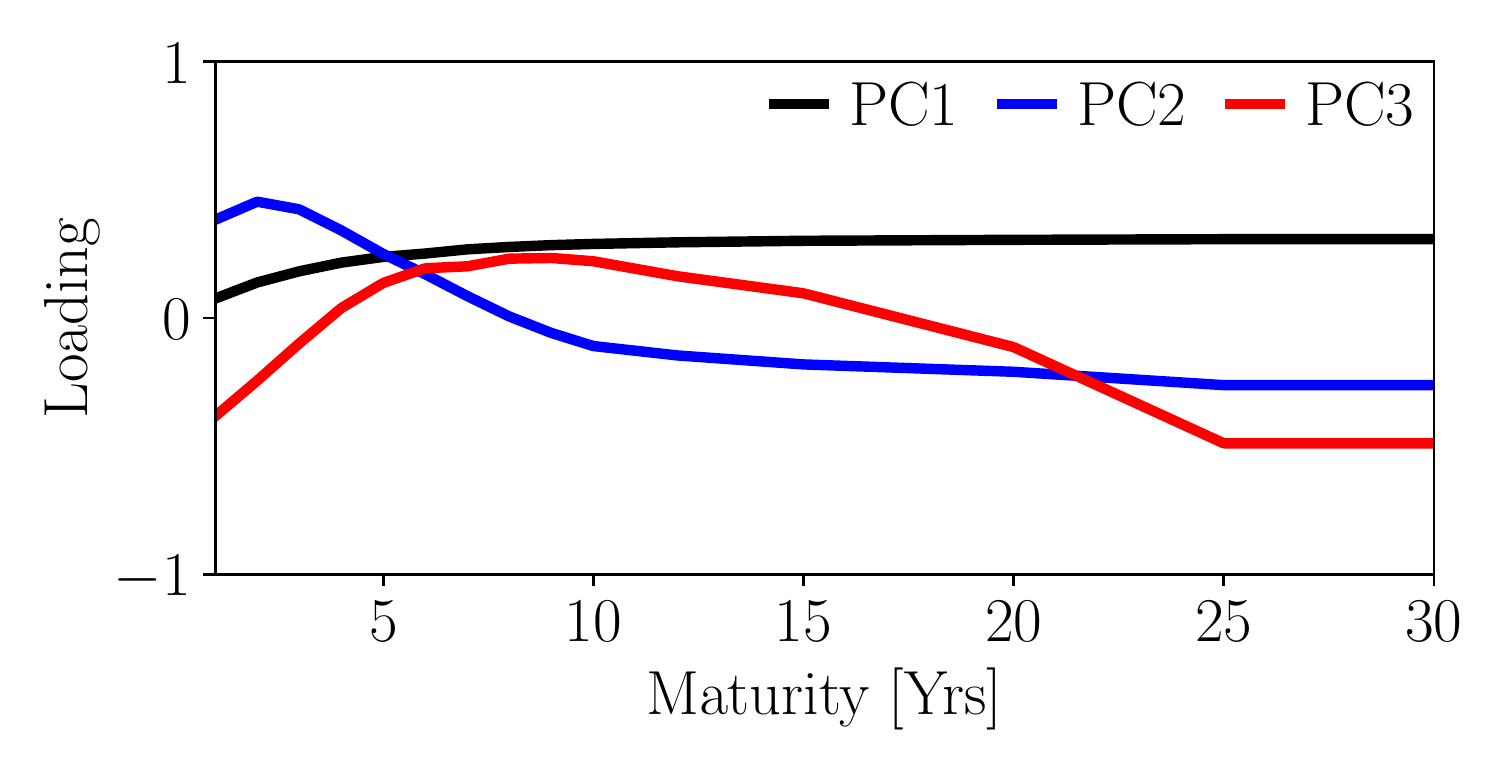}
		\caption{New Zealand (NZ)}    
	\end{subfigure}
	
	\begin{subfigure}[t]{0.5\textwidth}
		\centering
		\includegraphics[width=0.49\textwidth,trim={0 1.2cm 0 0},clip]{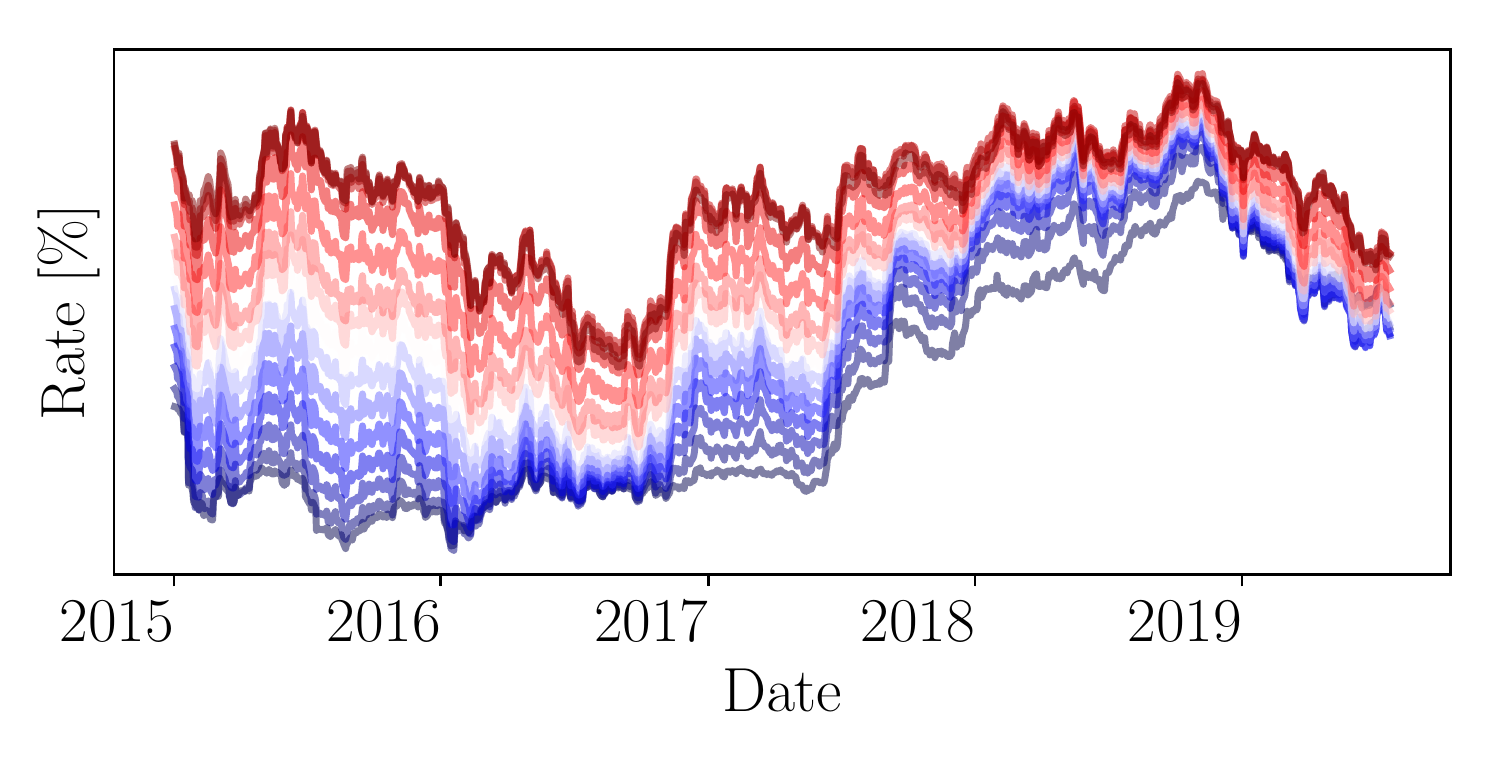}
		\includegraphics[width=0.49\textwidth,trim={0 1.2cm 0 0},clip]{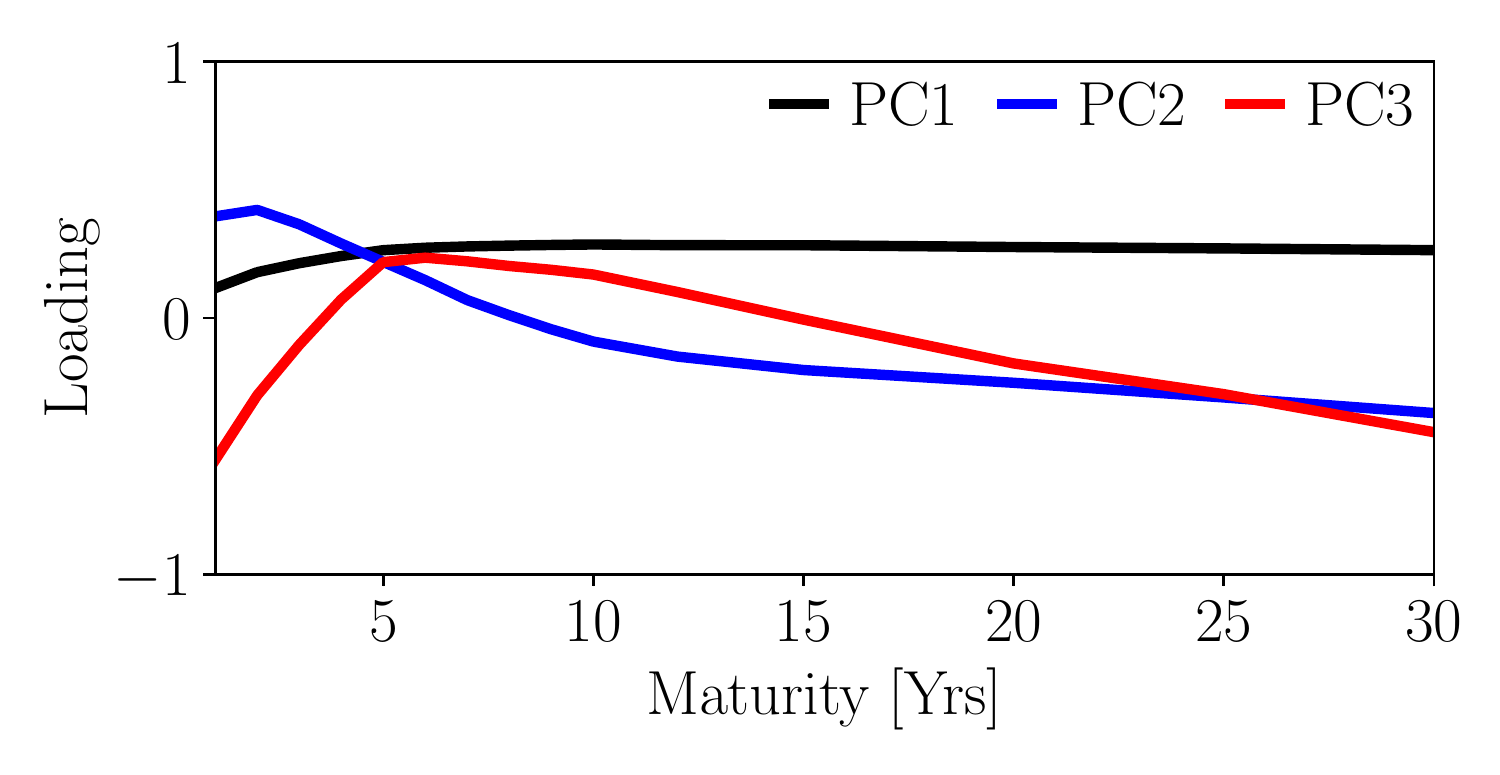}
		\caption{Canada (CA)}    
	\end{subfigure}
	
	\begin{subfigure}[t]{0.5\textwidth}
		\centering
		\includegraphics[width=0.49\textwidth,trim={0 1.2cm 0 0},clip]{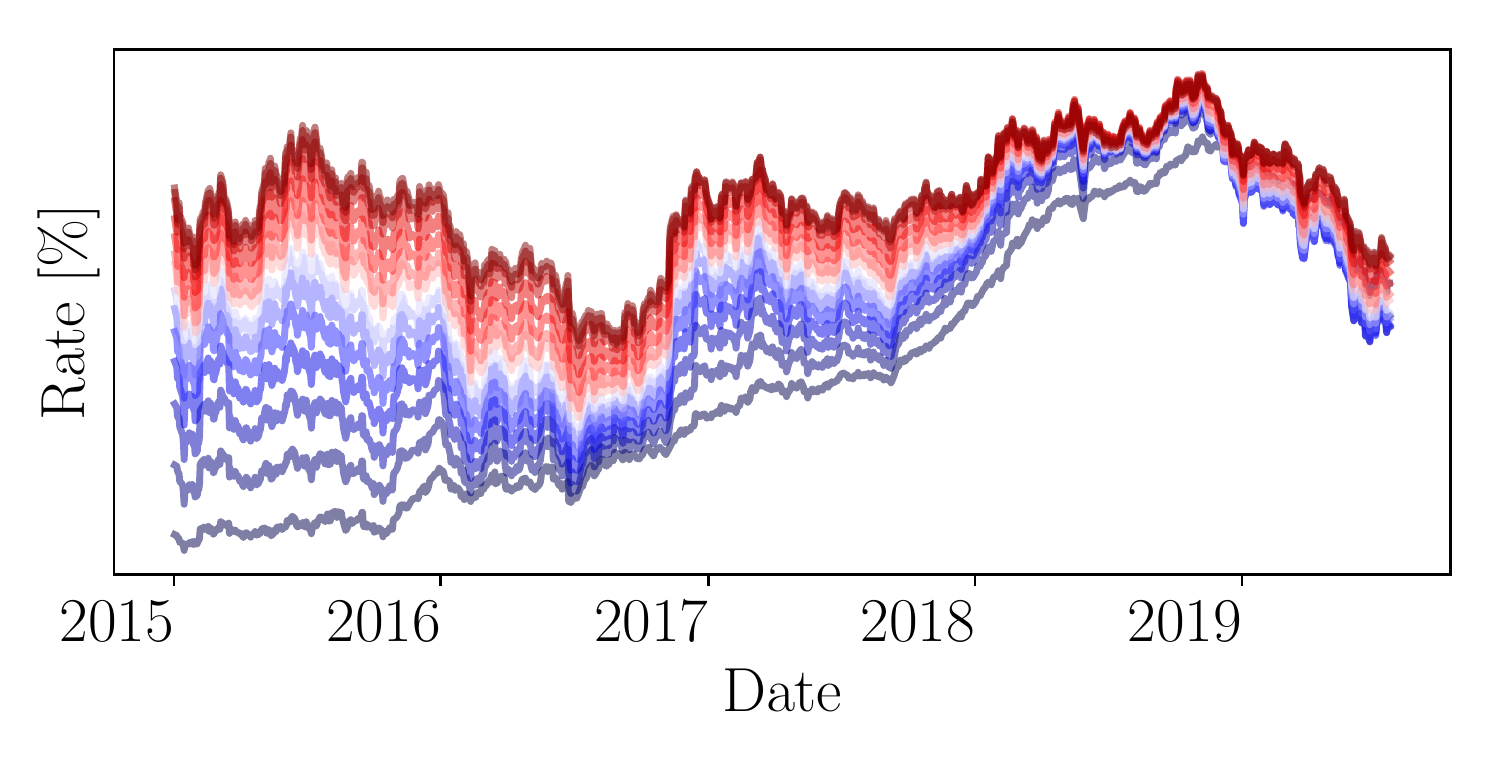}
		\includegraphics[width=0.49\textwidth,trim={0 1.2cm 0 0},clip]{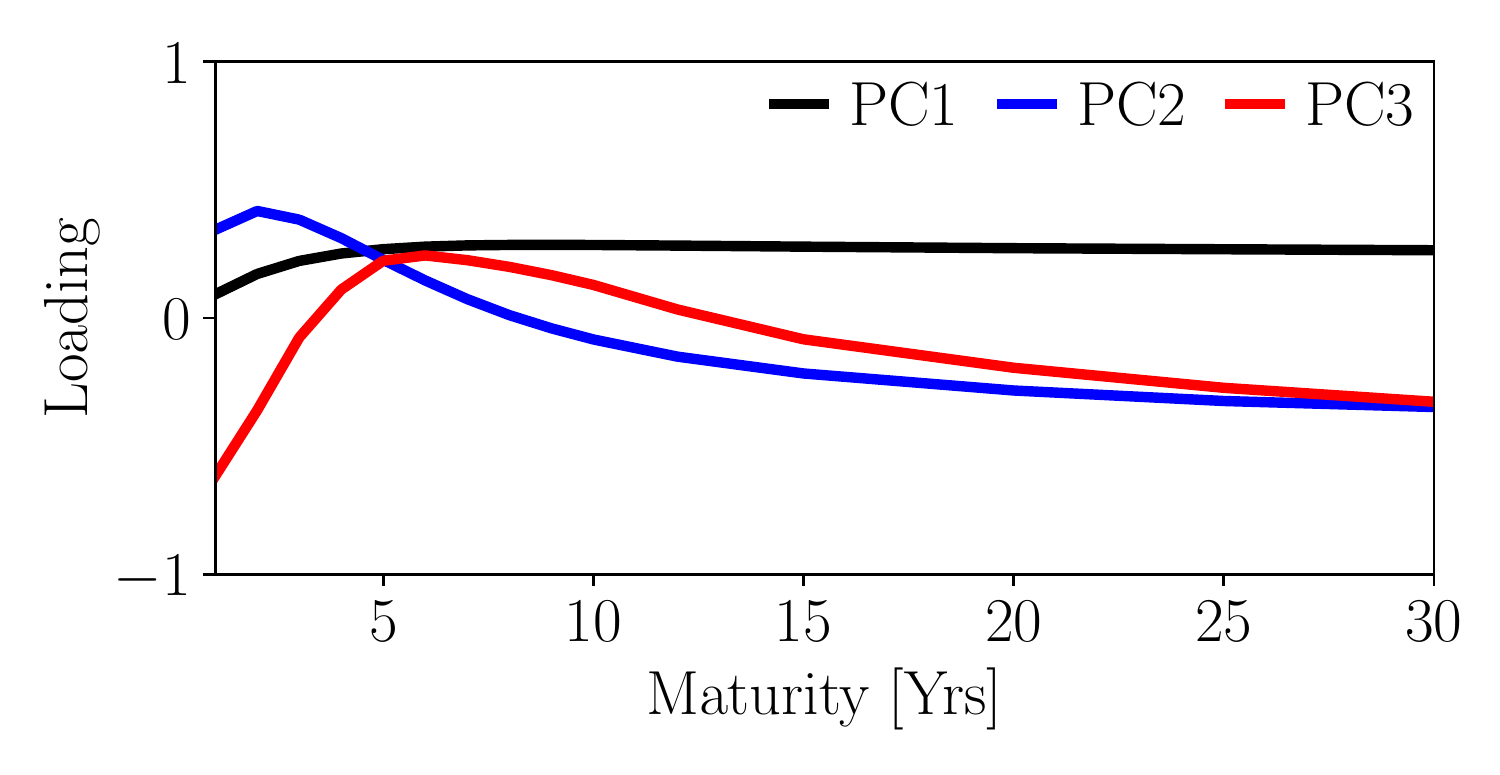}
		\caption{United States (US)}    
	\end{subfigure}

	\caption[]{\label{fig:domestic_PCA}Weekly swap rates\footnotemark \! for each economy with maturities $\{1,2,3,4,5,6,7,8,9,10,12,15,20,25,30\}$ years (respectively coloured from blue to red) during the period 2015-01-01 to 2019-07-01 (left panel) and their corresponding level, slope and curvature components obtained from the PCA of the swap weekly returns (right panel).}
\end{figure}

Firstly, as a complementary and preliminary assessment of the commonality of returns within different country IRS curves, we performed a principal component analysis for each of the eight economies independently, to obtain their dominant domestic principal components, that is, their domestic level, slope and curvature factors. The loadings of the three leading factors within each domestic IRS curve is shown in Figure \ref{fig:domestic_PCA}, and the percentage of variance explained by each component is reported in Table \ref{tab:domestic_PCA}. 

In agreement with the existing literature, all economies exhibit similar loadings across the three leading principal components. Moreover, the explanatory power of the components is consistent across all economies, whereby the first principal component (level) explains $\approx 90 \%$, and the second principal component (slope) explains $\approx 5 \%$, and the third principal component (curvature) explains $\approx 1\%$ of the variation in interest rate changes. We interpret this as indicating the existence of three leading dominant factors, that is the \textit{global level}, \textit{global slope} and \textit{global curvature} factors. With these preliminary and suggestive results, we now proceed to evaluate the common global risk factors with the proposed multilinear model.

\vspace{0.25cm}

\begin{center}
	\def\arraystretch{1.1}
	\begin{tabular}[h!]{ c || c | r | r} 
		\hline
		\textbf{Economy} & \textbf{Level} & \textbf{Slope} & \textbf{Curvature} \\
		\hline	\hline
		SF & $87.88$ & $10.02$ & $1.16$ \\
		EU & $94.15$ & $4.78$ & $0.66$ \\
		GB & $95.29$ & $3.83$ & $0.56$ \\
		JP & $82.04$ & $14.10$ & $2.28$ \\
		AU & $92.84$ & $4.94$ & $0.95$ \\
		NZ & $92.30$ & $5.76$ & $0.87$ \\
		CA & $93.14$ & $5.74$ & $0.66$ \\
		US & $95.30$ & $4.06$ & $0.47$ \\
		\hline\hline
	\end{tabular}
	\captionof{table}{Explanatory power [\%] of each principal component for the eight economies considered.} \label{tab:domestic_PCA} 
\end{center}

\vspace{0.25cm}

\newpage

\null

\footnotetext{Source: Bloomberg.}

\newpage

\subsection{Global analysis}

\label{section:analysis}

\begin{figure}[h]
	\captionsetup[subfigure]{aboveskip=-10pt,belowskip=0pt}
	\centering
	
	\vspace{-0.2cm}
	
	\begin{subfigure}[t]{0.45\textwidth}
		\centering
		\includegraphics[width=1\textwidth,trim={0 0.4cm 0 0},clip]{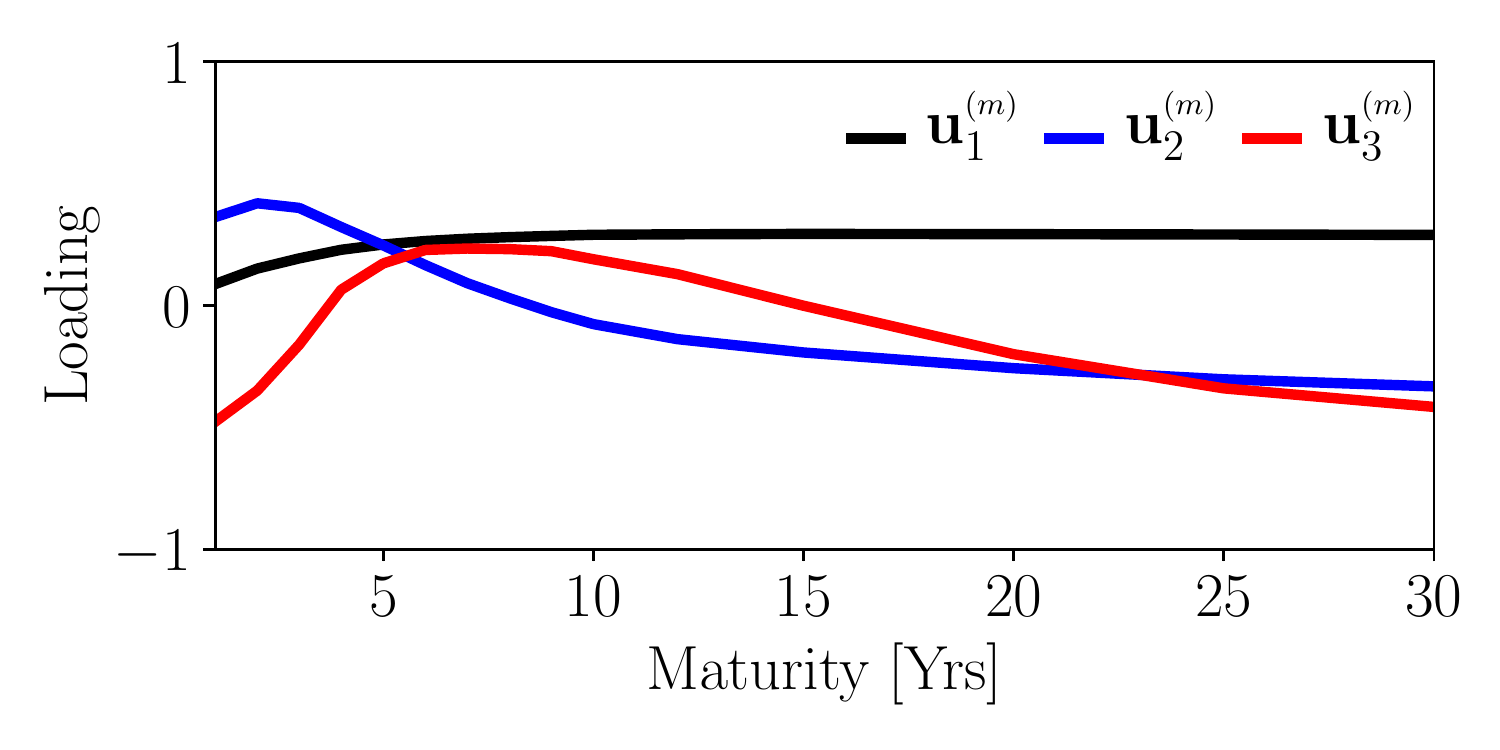} 
		\caption{\label{fig:global_loadings_maturity} Maturity-domain factor loadings.}    
	\end{subfigure}

	\begin{subfigure}[t]{0.45\textwidth}
		\centering
		\includegraphics[width=1\textwidth,trim={0 0.4cm 0 0},clip]{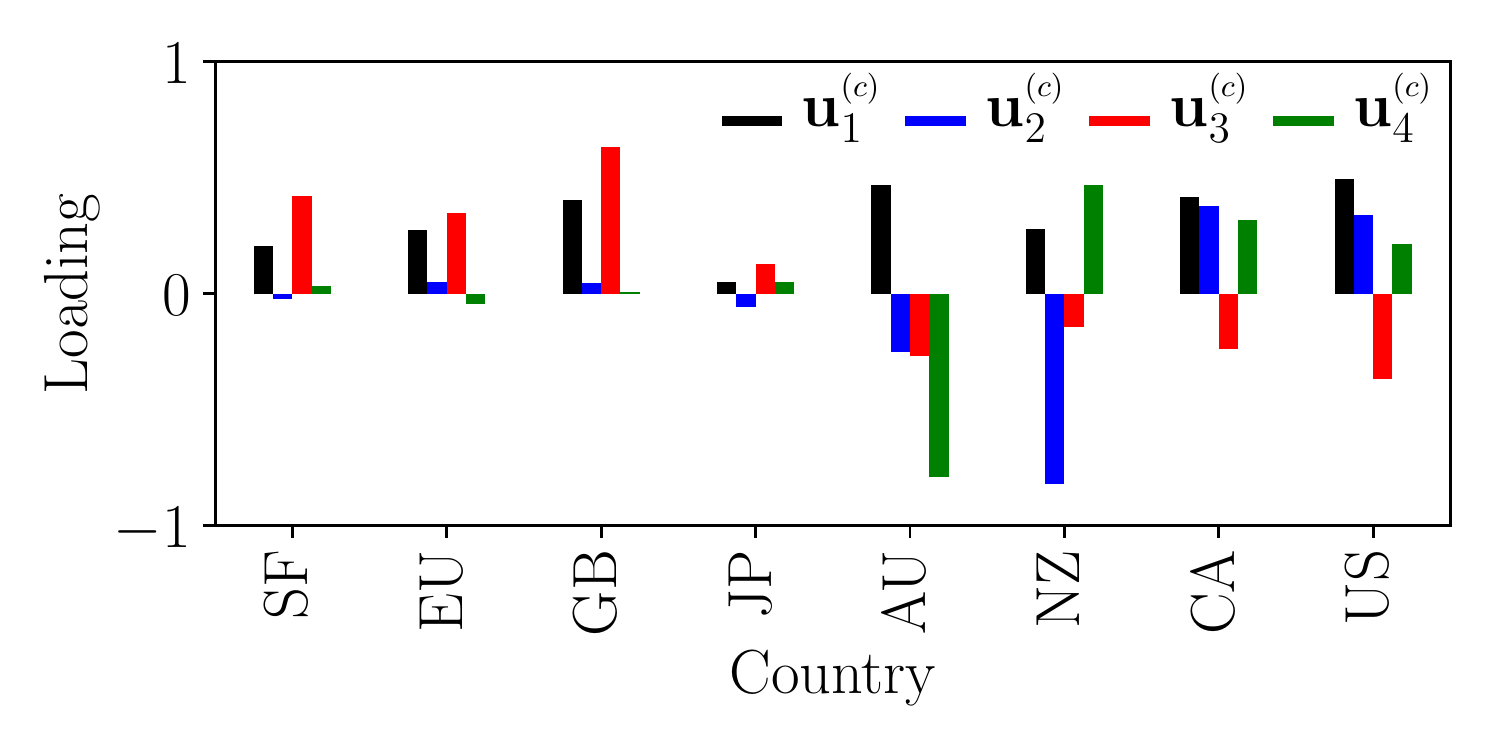}
		\caption{ \label{fig:global_loadings_country_1} Country-domain factor loadings (1--4).}    
	\end{subfigure}
	
		\begin{subfigure}[t]{0.45\textwidth}
		\centering
		\includegraphics[width=1\textwidth,trim={0 0.4cm 0 0},clip]{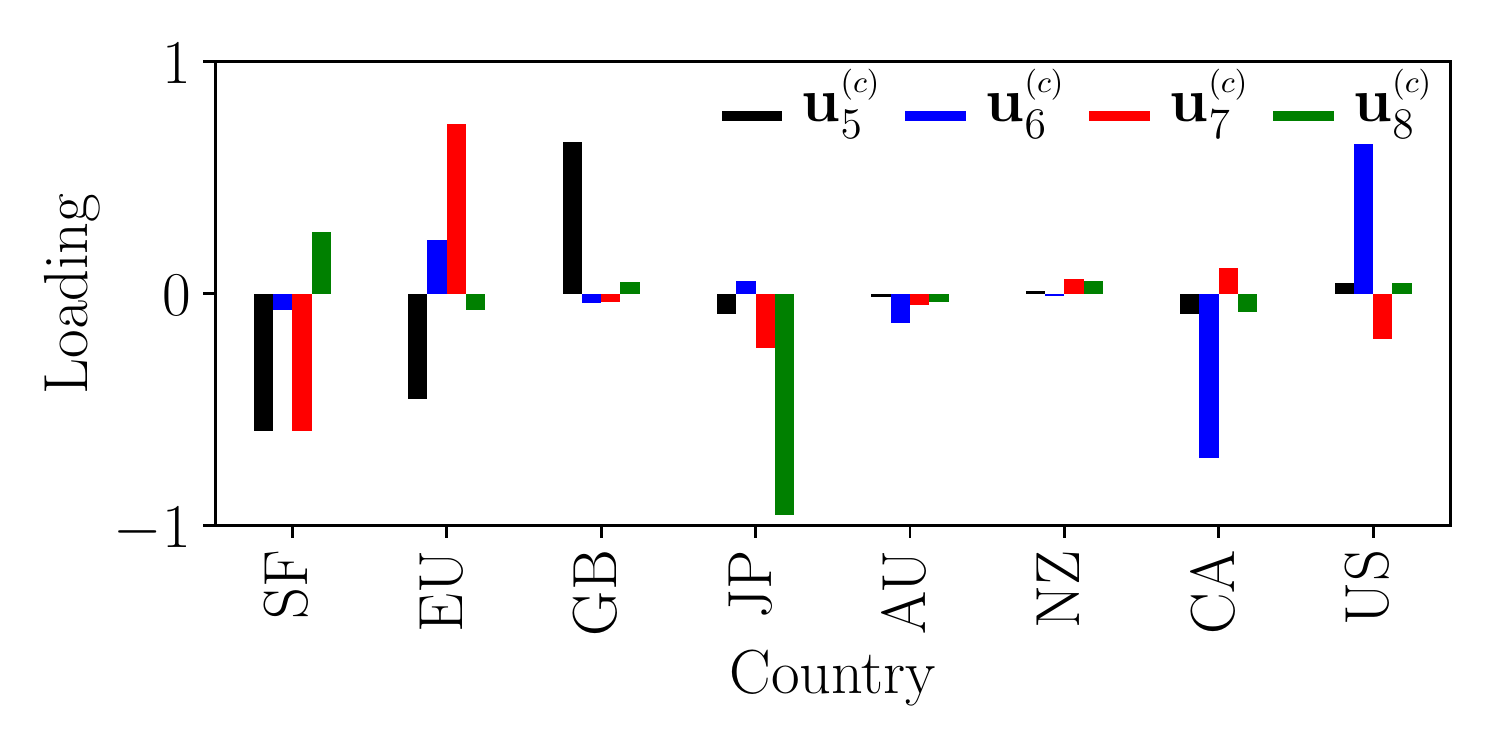} 
		\caption{\label{fig:global_loadings_country_2} Country-domain factor loadings (5--8).}    
	\end{subfigure}
	
	\caption{\label{fig:global_loadings} Loadings of the three leading maturity-domain global factors (top panel) and of the country-domain global factors (middle and bottom panels).}
\end{figure}

In this section we evaluate the results obtained from multilinear analysis of the international IRS dataset. The implementation procedure is summarised as follows\footnote{The data analysis was implemented using our own Python Higher-Order Tensor ToolBOX (HOTTBOX) \cite{Kisil2018}.}:
\begin{enumerate}[label=(\roman*)]
	\item The weekly IRS returns at the $t$-th week were tensorized to form the matrix-valued sample, $\X_{t} \in \domR^{I_{m} \times I_{c}}$, as described in (\ref{eq:tensorize});
	\item The parameters of the model ($\sigma^{2}$, $\boldTheta^{(m)}$, $\boldTheta^{(c)}$) were estimated using the analytic estimators in (\ref{eq:estimate_variance})-(\ref{eq:estimate_country});
	\item The global maturity-domain and country-domain factors, $\U^{(m)}$ and $\U^{(c)}$, and their associated eigenvalues, $\boldLambda^{(m)}$ and $\boldLambda^{(c)}$, were obtained from the eigendecompositions of $\boldTheta^{(m)}$, and $\boldTheta^{(c)}$, as shown in (\ref{eq:EVD_maturity})-(\ref{eq:EVD_country}).
\end{enumerate}
The loadings of the three leading maturity-domain factors, $\{ \u_{i}^{(m)} \}_{i=1}^{3}$, are plotted in Figure \ref{fig:global_loadings_maturity}, and their corresponding explanatory powers, $\{ \lambda_{i}^{(m)} \}_{i=1}^{3}$, are presented in Table \ref{tab:global_PCA_maturity}. The interpretation of the maturity-domain loadings is analogous to that of traditional domestic PCA. The maturity-domain factor loadings resemble the components obtained from domestic principal components (see Figure \ref{fig:domestic_PCA}), and therefore confirm the existence of a common set of bases shared by all economies. Furthermore, the explanatory powers of these factors are in line with that observed from the domestic analyses, which further strengthens the validity of our findings. The obtained maturity-domain factors clearly serve as a \textit{stencil} for describing the term structure within each domestic IRS curve, and as such we refer to these as the \textit{global level}, \textit{global slope} and \textit{global curvature}.

Additionally, the country-domain factors loadings, $\{ \u_{i}^{(c)} \}_{i=1}^{I_{c}}$, are visualized in Figures \ref{fig:global_loadings_country_1}--\ref{fig:global_loadings_country_2}, and their corresponding explanatory powers, $\{ \lambda_{i}^{(c)} \}_{i=1}^{I_{c}}$, are presented in Table \ref{tab:global_PCA_country}. The most dominant factor, $\u_{1}^{(c)}$, has positive loadings across all economies, and can be thought of as the \textit{global risk premium}, analogous to the level factor in the maturity-domain. Notice that this factor also explains a significant portion of the international IRS variance. The remaining country-domain factors represent interpretable macroeconomic factors concerning subsets of the considered economies. These results demonstrate the direct applicability of the proposed approach for gaining physical insight into the global macroeconomic environment in a straightforward and compact manner, owing to the small number of parameters required to fully describe the global fixed income universe.

With reference to the previous section, the maturity-domain (listed in Table \ref{tab:global_PCA_maturity}) and country-domain (listed in Table \ref{tab:global_PCA_country}) factor loadings can be directly employed for global macroeconomic hedging and risk management. We conclude this section by reiterating the practical advantage of the multilinear framework, namely: (i) the reduction in parameters required to optimize the global portfolio, which for $I_{m}=15$ and $I_{c}=8$ reduces from $I_{m}I_{c}=120$ to $(I_{m}+I_{c}) = 23$ portfolio weight parameters; and (ii) the parsimonious description of the global risk in terms of parallel maturity-domain and country-domain risk factors that can facilitate the investor's decision making process.

\newpage

\vspace{-0.25cm}

\begin{center}
	\def\arraystretch{1}
	\begin{tabular}[h]{ c || c | l | r} 
		\hline
		\textbf{Factor} & \textbf{Symbol} &  \makecell[l]{\textbf{Economic} \\ \textbf{interpretation}} & \makecell[l]{\textbf{Variance} \\ \textbf{explained} [\%]} \\
		\hline\hline
		1 & $\u_{1}^{(m)}$ & Global level & $92.37$ \\\hline
		2 & $\u_{2}^{(m)}$ & Global slope & $5.90$  \\\hline
		3 & $\u_{3}^{(m)}$ & Global curvature & $0.97$ \\\hline\hline
	\end{tabular}
	\captionof{table}{Economic interpretation and explanatory power of the three leading global factors in the maturity-domain.} \label{tab:global_PCA_maturity} 
\end{center}

\vspace{-0.4cm}

\begin{center}
	\def\arraystretch{1}
	\begin{tabular}[h]{ c || c | l | r } 
		\hline
		\textbf{Factor} & \textbf{Symbol} & \makecell[l]{\textbf{Economic} \\ \textbf{interpretation}} & \makecell[l]{\textbf{Variance} \\ \textbf{explained} [\%]} \\
		\hline\hline
		1 & $\u_{1}^{(c)}$ & Global risk premium & $71.62$ \\ \hline
		2 & $\u_{2}^{(c)}$ & \makecell[l]{(AU, NZ) vs. rest} & $8.34$  \\\hline
		3 & $\u_{3}^{(c)}$ & \makecell[l]{(SF, EU, GB, JP) vs.\\ (AU, NZ, CA, US)} & $5.72$ \\\hline
		4 & $\u_{4}^{(c)}$ & \makecell[l]{AU vs. rest} & $4.21$ \\\hline
		5 & $\u_{5}^{(c)}$ & GB vs (SF, EU) & $3.46$ \\\hline
		6 & $\u_{6}^{(c)}$ & CA vs US & $3.02$ \\\hline
		7 & $\u_{7}^{(c)}$ & SF vs EU & $2.13$ \\\hline
		8 & $\u_{8}^{(c)}$ & JPY & $1.5$ \\\hline\hline
	\end{tabular}
	\captionof{table}{Economic interpretation and explanatory power of the eight leading global factors in the country-domain.} \label{tab:global_PCA_country} 
\end{center}

\pagebreak

\section{Conclusions}

A unifying tensor-valued framework for modelling the global risk factors shared by multiple domestic term structures has been introduced. By virtue of the multilinear approach (as opposed to the current ``flat-view'' multivariate ones), the proposed approach has been shown to decompose the overall multivariate covariance structure of international asset returns into maturity-domain covariance and country-domain covariance. In this way, the proposed analysis: (i) achieves a significant reduction in the number of parameters required to fully describe the international investment universe; and (ii) offers a physically interpretable setting for estimating and identifying global risk factors. As a natural extension of the proposed framework, we have derived analytic solutions to global hedging and portfolio management, which allows the investor to gain enhanced control over the portfolio risk within two independent domains -- maturity and country. An empirical analysis has been performed on the interest rate swaps curves for eight developed economies, and the results have confirmed the existence of common global risk factors. The results are supported by our own Python toolbox for tensor analysis \cite{Kisil2018}. Although we have focused the analysis on fixed income assets, the methodology can be generalised to any asset class.  



\section*{Acknowledgments}

The author would like to express his sincere gratitude to Danilo P. Mandic, Vladimir Lucic and Anoosh Lachin for their very constructive comments.



\footnotesize

\bibliographystyle{IEEEtran}
\bibliography{./Bibliography} 

\begin{thebibliography}{10}
\providecommand{\url}[1]{#1}
\csname url@samestyle\endcsname
\providecommand{\newblock}{\relax}
\providecommand{\bibinfo}[2]{#2}
\providecommand{\BIBentrySTDinterwordspacing}{\spaceskip=0pt\relax}
\providecommand{\BIBentryALTinterwordstretchfactor}{4}
\providecommand{\BIBentryALTinterwordspacing}{\spaceskip=\fontdimen2\font plus
\BIBentryALTinterwordstretchfactor\fontdimen3\font minus
  \fontdimen4\font\relax}
\providecommand{\BIBforeignlanguage}[2]{{%
\expandafter\ifx\csname l@#1\endcsname\relax
\typeout{** WARNING: IEEEtran.bst: No hyphenation pattern has been}%
\typeout{** loaded for the language `#1'. Using the pattern for}%
\typeout{** the default language instead.}%
\else
\language=\csname l@#1\endcsname
\fi
#2}}
\providecommand{\BIBdecl}{\relax}
\BIBdecl

\bibitem{Litterman1991}
R.~Litterman and J.~Scheinkmann, ``{Common Factors Affecting Bond Returns.}''
  \emph{{Journal of Fixed Income}}, vol.~1, pp. 54--61, 1991.

\bibitem{Jolliffe1986}
I.~T. Jolliffe, \emph{{Principal Component Analysis}}.\hskip 1em plus 0.5em
  minus 0.4em\relax New York: Springer--Verlag, 1986.

\bibitem{Rodrigues1997}
A.~P. Rodrigues, ``{Term Structure and Volatility Shocks},'' \emph{{Working
  Paper, Federal Reserve Bank of New York}}, 1997.

\bibitem{Driessen2003}
J.~Driessen, B.~Melenberg, and T.~Nijman, ``{Common Factors in International
  Bond Returns},'' \emph{{Journal of International Money and Finance}},
  vol.~22, pp. 629--656, 2003.

\bibitem{Novosyolov2008}
A.~Novosyolov and D.~Satchkov, ``{Global Term Structure Modeling using
  Principal Components Analysis},'' \emph{{Journal of Asset Management}},
  vol.~9, pp. 49--60, 2008.

\bibitem{Flury1988}
B.~Flury, \emph{{Common Principal Components and Related Multivariate
  Models}}.\hskip 1em plus 0.5em minus 0.4em\relax New York: Wiley, 1988.

\bibitem{Juneja2012}
J.~Juneja, ``{Common Factors, Principal Components Analysis, and the Term
  Structure of Interest Rates},'' \emph{{International Review of Financial
  Analysis}}, vol.~24, pp. 48--56, 2012.

\bibitem{Tucker1958}
L.~R. Tucker, ``{An Inter-Battery Method of Factor Analysis},''
  \emph{{Psychometrika}}, vol.~23, pp. 111--136, 1958.

\bibitem{Perignon2007}
C.~P\'erignon, D.~R. Smith, and C.~Villa, ``{Why Common Factors in
  International Bond Returns Are Not so Common},'' \emph{{Journal of
  International Money and Finance}}, vol.~26, pp. 284--304, 2007.

\bibitem{Kolda2009}
T.~G. Kolda and B.~W. Bader, ``{Tensor Decompositions and Applications},''
  \emph{{SIAM Review}}, vol.~51, no.~3, pp. 455--500, 2009.

\bibitem{Mandic2015_3}
A.~Cichocki, D.~P. Mandic, A.~H. Phan, C.~F. Caiafa, G.~Zhou, Q.~Zhao, and
  L.~De~Lathauwer, ``{Tensor Decompositions for Signal Processing
  Applications},'' \emph{{IEEE Signal Processing Magazine}}, vol. 145, pp.
  145--163, 2015.

\bibitem{Mandic2017_2}
A.~Cichocki, A.~H. Phan, Q.~Zhao, N.~Lee, I.~Oseledets, and D.~P. Mandic,
  ``{Tensor Networks for Dimensionality Reduction and Large-Scale
  Optimizations. Part 1: Low--Rank Tensor Decompositions},'' \emph{{Foundations
  and Trends in Machine Learning}}, vol.~9, no. 4--5, pp. 249--429, 2017.

\bibitem{Mandic2017_3}
A.~Cichocki, A.~H. Phan, Q.~Zhao, N.~Lee, I.~Oseledets, M.~Sugiyama, and D.~P.
  Mandic, ``{Tensor Networks for Dimensionality Reduction and Large-Scale
  Optimizations. Part 2: Applications and Future Perspectives},''
  \emph{{Foundations and Trends in Machine Learning}}, vol.~9, no.~6, pp.
  431--673, 2017.

\bibitem{Sidiropoulos2017_1}
N.~D. Siridopoulos, L.~De~Lathauwer, X.~Fu, K.~Huang, E.~E. Papalexakis, and
  C.~Faloutsos, ``{Tensor Decomposition for Signal Processing and Machine
  Learning},'' \emph{{IEEE Transactions on Signal Processing}}, vol.~65,
  no.~13, pp. 3551--3582, 2017.

\bibitem{Hoff2011}
P.~D. Hoff, ``{Separable Covariance Arrays via the Tucker Product, with
  Applications to Multivariate Relational Data},'' \emph{{Bayesian Analysis}},
  vol.~6, no.~2, pp. 179--196, 2011.

\bibitem{Scalzo2019_1}
B.~Scalzo~Dees and D.~P. Mandic, ``{A Statistically Identifiable Model for
  Tensor-Valued Gaussian Random Variables},'' \emph{{arXiv:1911.02915}}, 2019.

\bibitem{Magnus1985}
J.~R. Magnus and H.~Neudecker, ``{Matrix Differential Calculus with
  Applications to Simple, Hadamard, and Kronecker Products},'' \emph{{Journal
  of Mathematical Psychology}}, vol.~29, pp. 474--492, 1985.

\bibitem{Tucker1966}
L.~R. Tucker, ``{Some Mathematical Notes on Three-Mode Factor Analysis},''
  \emph{{Psychometrika}}, vol.~31, no.~3, pp. 279--311, 1966.

\bibitem{DeLathauwer2000}
L.~De~Lathauwer, B.~D. Moor, and J.~Vandewalle, ``{A Multilinear Singular Value
  Decomposition},'' \emph{{SIAM Journal on Matrix Analysis and Applications}},
  vol.~21, no.~4, pp. 1253--1278, 2000.

\bibitem{Weiss2005}
N.~A. Weiss, P.~T. Holmes, and M.~Hardy, \emph{{A Course in
  Probability}}.\hskip 1em plus 0.5em minus 0.4em\relax Pearson Addison Wesley,
  2005.

\bibitem{Solomon2000}
``{Principles of Principal Components: A Fresh Look at Risk, Hedging, and
  Relative Value},'' \emph{{Research Report, Solomon Smith Barney}}, 2000.

\bibitem{CreditSuisse2012}
``{PCA Unleashed},'' \emph{{Research Report, Credit Suisse}}, 2015.

\bibitem{StandardChartered2013}
``{Introducing a Relative Value Tool for Swaps},'' \emph{{Research Report,
  Standard Chartered}}, 2013.

\bibitem{TDSecurities2015}
``{Market Musings -- Relative Value Across the U.S. Swap Surface: A PCA
  Approach},'' \emph{{Research Report, TD Securities}}, 2015.

\bibitem{Kisil2018}
I.~Kisil, B.~Scalzo~Dees, A.~Moniri, G.~G. Calvi, and D.~P. Mandic, ``{HOTTBOX:
  Higher Order Tensor ToolBOX},'' https://hottbox.github.io.

\end{thebibliography}

\end{document}